\documentclass[aps,amssymb,pre, amsmath, floatfix, preprint]{revtex4}
\usepackage{graphics}
\usepackage{psfrag}
\usepackage{epsfig}
\usepackage{subfigure}
\begin{document}
\bibliographystyle{prsty}

\title{Long Range Correlations in Granular Shear Flow II: \\  Theoretical Implications}
\author{Gregg Lois$^{(1)}$}
\author{Ana\"el Lema\^{\i}tre$^{(2)}$}
\author{Jean M. Carlson$^{(1)}$}
\affiliation{
$^{(1)}$ Department of physics, University of California, Santa Barbara, California 93106, U.S.A.}
\affiliation{$^{(2)}$ Institut Navier-- LMSGC, 2 all\'ee K\'epler, 77420 Champs-sur-Marne, France} 
\date{\today}

\begin{abstract}
Numerical simulations are used to test the kinetic theory constitutive relations of inertial granular shear flow.  These predictions are shown to be accurate in the dilute regime, where only binary collisions are relevant, but underestimate the measured value in the dense regime, where force networks of size $\xi$ are present.  
The discrepancy in the dense regime is due to non-collisional forces that we measure directly in our simulations and arise from elastic deformations of the force networks.  We model the non-collisional stress by summing over all paths that elastic waves travel through force networks.  This results in an analytical theory that successfully predicts the stress tensor over the entire inertial regime without any adjustable parameters.  
\end{abstract}
\maketitle

\section{Introduction}
Granular materials exhibit a broad spectrum of behaviors that have been difficult to capture theoretically, especially in the dense regime.  A fundamental question is when, and whether, collective motion becomes important for understanding the macroscopic state of the system~\cite{aransonreview}.  This is a particularly complex issue for granular flows, where the material structure remains amorphous on all scales.  While a noticeable structural change would surely help pinpoint degrees of freedom that govern collective dynamics, granular flows require us to use the dynamics to search for structure.

A central tool in this process is numerical simulation, and new insights from simulations can provide important clues for theory.  In the companion paper~\cite{gajcompanion1} we have presented evidence for a growing length scale related to correlations between grain forces.  These correlations arise from force chain networks that span the space between grains and transmit forces elastically.  The emergence of force networks alters the mechanisms of momentum transfer, and contact force statistics depend on the size $\xi$ of force networks.     

Since contact forces are sensitive to the value of $\xi$, the derivation of constitutive relations to describe the stress tensor must also be based on properties of the networks.  While network properties have been incorporated into previous studies of static granular packings~\cite{qmodel, socolaralpha, claudinstress,nicodemi,socolarforce, ottoforce, bouchaudforce}, the importance of force networks in the inertial regime has not been adequately taken into account.  Instead, constitutive relations are generally obtained using comparisons with liquids~\cite{ogawa,haff} and especially kinetic theory~\cite{jenkinssavage,lun,jenkins}, which can be extended to treat dense materials~\cite{garzodufty,lutsko1,lutsko2} as long as correlations beyond nearest neighbors are absent.

In this paper we examine two models for constitutive relations in granular flow:  kinetic theory and a new model, referred to as the force network model.  Kinetic theory is rooted in the dilute limit and assumes no long range correlations between grains.  The force network model is motivated by our simulations and attempts to capture the effects of finite sized force networks in a mean-field framework, which spans system behavior from dilute to dense regimes.  

The predictions of these models are compared using measurements of the stress tensor $\Sigma_{\alpha\beta}$, which is given by
\begin{equation}
\Sigma_{\alpha\beta} V = \sum_{i=1}^N  m^i (v_\alpha^i-u_\alpha) (v_\beta^i-u_\beta) +  \sum_{\{i,j\}=1}^{c}  \mbox{\boldmath $\sigma$}^{ij}_\alpha F^{ij}_\beta.
\label{stresstensor} 
\end{equation}
In this equation, Greek subscripts denote components, Roman superscripts denote grains, and $V$ represents the volume.
The first term quantifies stress resulting from fluctuating velocities, which is related to the mass of each grain $m^i$ and the difference between grain velocity ${\bf v}^i$ and average velocity ${\bf u}$ (where bold face symbols denote full, three component vectors).  The second term arises from contacts between grains: it depends on the contact forces ${\bf F}^{ij}$ and the distance between contacting grains $\mbox{\boldmath $\sigma$}^{ij}$.  The sum is taken over all contacts $\{i,j\}$ in the system.

We focus on the second term, called the static stress, since it is directly associated with contact forces, and we specialize to non-frictional systems where $\mbox{\boldmath $\sigma$}^{ij} \times {\bf F}^{ij} = 0$.  This allows us to write the static stress tensor $\Sigma^{\mathrm{s}}_{\alpha\beta}$ as
\begin{equation}
%\Sigma^{\mathrm{s}}_{\alpha\beta} V = \frac{1}{2} \sum_{\{i,j\}=1}^{c}  \hat{\sigma}^{ij}_\alpha \hat{\sigma}^{ij}_\beta \sigma^{ij} F^{ij},
\Sigma^{\mathrm{s}}_{\alpha\beta} V =  \sum_{\{i,j\}=1}^{c}  \hat{\sigma}^{ij}_\alpha \hat{\sigma}^{ij}_\beta \sigma^{ij} F^{ij},
\label{sstress}
\end{equation}
where $\mbox{\boldmath $\sigma$}^{ij} = \sigma^{ij} \hat{\sigma}^{ij}$ and $F^{ij}$ is the magnitude of the contact force between pairs $\{i,j\}$ of contacting grains.  

We begin in Section~\ref{kineticsection} with an investigation of dilute inertial flows and test the predictions of hard-sphere kinetic theory.  In Section~\ref{forcenetsection} we focus on dense inertial flows and introduce the force network model.

\section{Dilute Inertial Flows}
\label{kineticsection}
Over the past twenty five years~\cite{campbellreview, goldhirschreview} the kinetic theory of dense gases has been generalized to include granular flows, where thermal fluctuations are absent and energy is dissipated at each contact.  The dissipation mechanism most often considered is instantaneous collisions with constant restitution coefficient-- this is called hard-sphere kinetic theory~\cite{vannoijebook} .  In order to make progress using hard-sphere kinetic theory, it is necessary to {\em begin} by postulating that only binary collisions occur between grains.  Without this assumption, calculations quickly become intractable since high order correlations must be included in kinetic integrals.

The principal microscopic input to hard-sphere kinetic theory is the collision rule between grains. This relates the initial velocities of two interacting grains $\{ {\bf v'}^i$, ${\bf v'}^j \}$ to their final velocities  $\{ {\bf v}^i$, ${\bf v}^j \}$: 
\begin{equation}
({\bf v}^j - {\bf v}^i) \cdot \hat{\sigma}^{ij} = -e ({\bf v'}^j - {\bf v'}^i) \cdot \hat{\sigma}^{ij}.
\label{binarycollisions}
\end{equation}
The normal coefficient of restitution $e$ that appears in this equation determines the energy dissipated in each collision: for $e=1$ the system is elastic and no energy is dissipated; as $e$ is reduced to zero the energy dissipation scales as $1-e^2$.  

%Using the collision rule in Equation~(\ref{binarycollisions}), combined with the assumption that the collisions occur instantaneously and momentum is conserved, we can derive the magnitude of the impulse $I_{\mathrm{bc}}^{ij}$ between colliding grains.  This is given by
%\begin{equation}
%I^{ij}_{\mathrm{bc}} = (1+e) \mu^{ij} \left[ ({\bf v'}^j - {\bf v'}^i) \cdot \hat{\sigma}^{ij} \right]
%\label{binaryimpulse}
%\end{equation}
%and directed normal to the contact.  We label this impulse with ``bc'' since it is derived from the binary collision assumption; its value depends on the reduced mass $\mu$ of the interacting grains and the pre-collisional velocities ${\bf v'}$.

%\subsubsection{Quantifying the Dilute Regime}
Kinetic theory relies on the assumption that only binary collision are relevant and is therefore expected to break down as the density of the flow increases and long-lasting contacts arise~\cite{azanza, zhang1, shen, zhang2}.  
Quantitative bounds over which the binary collision assumption holds have only recently been estimated~\cite{gaj2, gaj3}.
The Contact Dynamics (CD) algorithm used here to simulate granular flows is well suited for testing the relevance of the binary collision assumption and bounding the dilute regime.  Like hard-sphere kinetic theory, the CD algorithm employs a normal coefficient of restitution.
%and assumes that collisions occur instantaneously.  
However the CD algorithm does not assume {\em a priori} that only binary collisions occur. 

On the contrary, we observe that multi-grain contacts are the dominant interaction in dense flows.  This is evident in force correlation measurements~\cite{gajcompanion1}, which identify a growing correlation length $\xi$.  This length scale should be an indicator of the breakdown of the binary collision assumption since it implies that grain forces are correlated over large distances and do not simply depend on nearest neighbors.  To see how this comes about, it is useful to measure the static stress tensor.      

Equation~(\ref{sstress}) gives the microscopic expression for the static stress tensor, which depends on contact forces between grains.  In the case that only binary collisions are considered, as in kinetic theory, the value of the static stress tensor is determined by inserting the binary collisional force into Equation~(\ref{sstress}).  The collisional force is the force that occurs between a pair of colliding grains that are not part of a force network.  Given the initial velocities ${\bf v'}^i$ of grains $i$, the collisional force between two grains is given by
\begin{equation}
F^{ij}_{\mathrm{bc}} = (1+e) \mu^{ij} \left[ ({\bf v'}^j - {\bf v'}^i) \cdot \hat{\sigma}^{ij} \right]/dt,
\label{binaryforce}
\end{equation}
where $e$ is the normal restitution coefficient, $\mu = m^i m^j/(m^i+m^j)$ is the reduced mass, and $\mbox{\boldmath $\hat{\sigma}$}^{ij}$ is the unit vector connecting the centers of grains $i$ and $j$.  This expression is equal to the change of momentum of grains $i$ and $j$, per simulation time step $dt$, due to binary interactions.  Since all of the parameters can be measured in simulations, the collisional force is a useful probe of the dynamics and was used in the companion paper to exhibit the presence of force networks. 

Inserting the collisional force into Equation~(\ref{sstress})
yields the ``collisional'' stress tensor
\begin{equation}
%\Sigma^{\mathrm{bc}}_{\alpha\beta} = \frac{1}{2} \sum_{\{i,j\}=0}^{c}  \hat{\sigma}^{ij}_\alpha \hat{\sigma}^{ij}_\beta \sigma^{ij} F_{\mathrm{bc}}^{ij},
\Sigma^{\mathrm{bc}}_{\alpha\beta} = \sum_{\{i,j\}=0}^{c}  \hat{\sigma}^{ij}_\alpha \hat{\sigma}^{ij}_\beta \sigma^{ij} F_{\mathrm{bc}}^{ij},
\label{bcstress}
\end{equation}
which is the stress tensor 
that all hard-sphere kinetic theories attempt to model~\cite{gaj3}.  This is because only binary collisions between grains are considered and contact forces can be no larger than the value given in Equation~(\ref{binaryforce}).  

If relevant interactions occur only via binary collisions, the collisional stress in Equation~(\ref{bcstress}) and the static stress in Equation~(\ref{sstress}) are identical.   However, a discrepancy between the collisional and static stress tensors indicates that momentum transfer is not carried out by binary collisions alone.
%n the presence of force networks, the average contact force between any two grains is greater than what is expected from binary collisions alone.  This is because the presence of additional contacts .  In this case, the static stress tensor is larger than the collisional stress tensor and kinetic theory is not applicable.  
Comparing the static and collisional stress tensors in Equations~(\ref{sstress}) and (\ref{bcstress}) therefore provides an opportunity to test the binary collision assumption and thereby determine when kinetic theory can be applied to hard-sphere granular flows.

\begin{figure}[htbp]
\begin{center}

\mbox{
\subfigure{\scalebox{.35}{\includegraphics{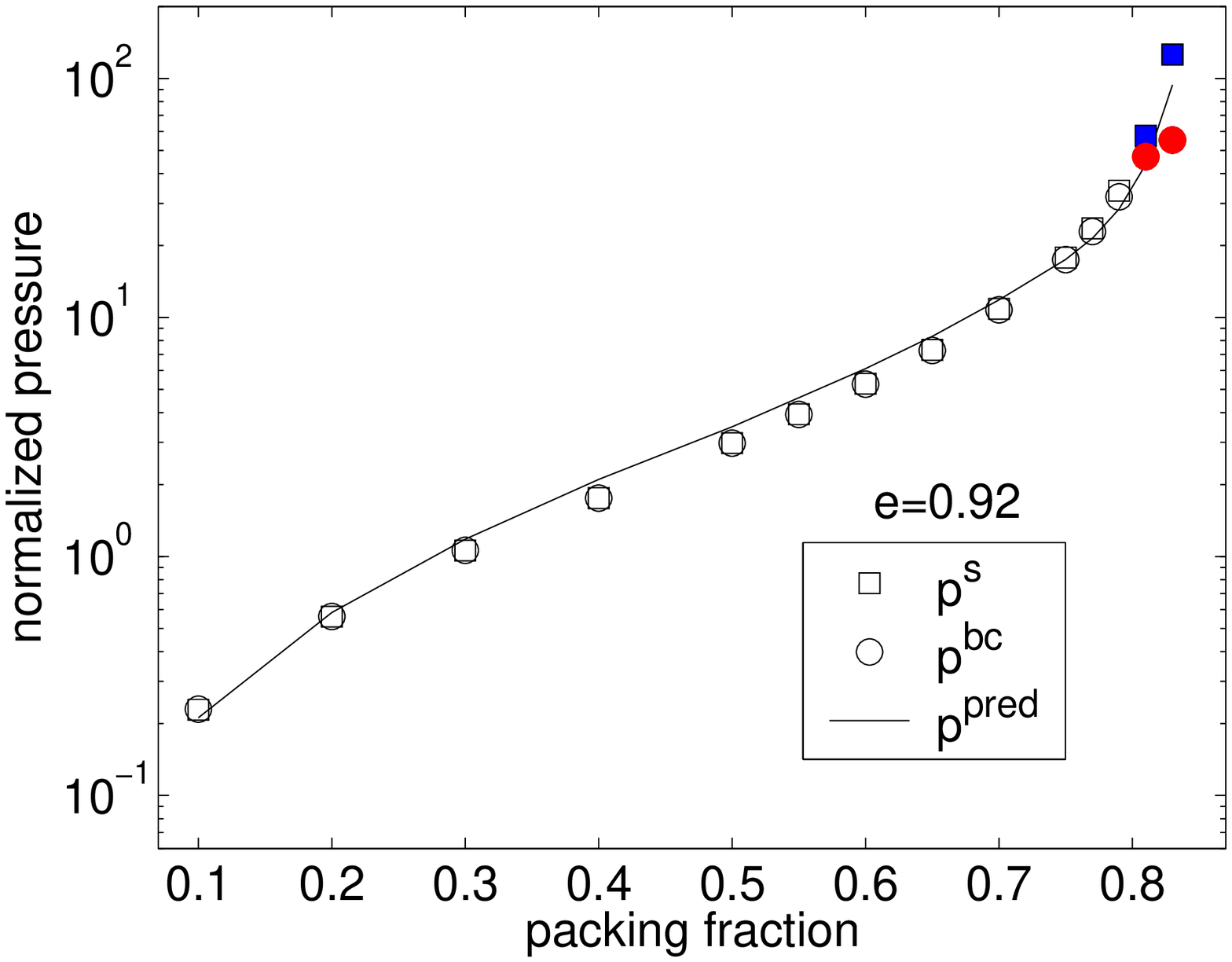}}}
\quad \quad
\psfrag{iyl}{\Huge{$C_v$}}
%\psfrag{ixl}{\Large{packing fraction}}
\subfigure{\scalebox{.35}{\includegraphics{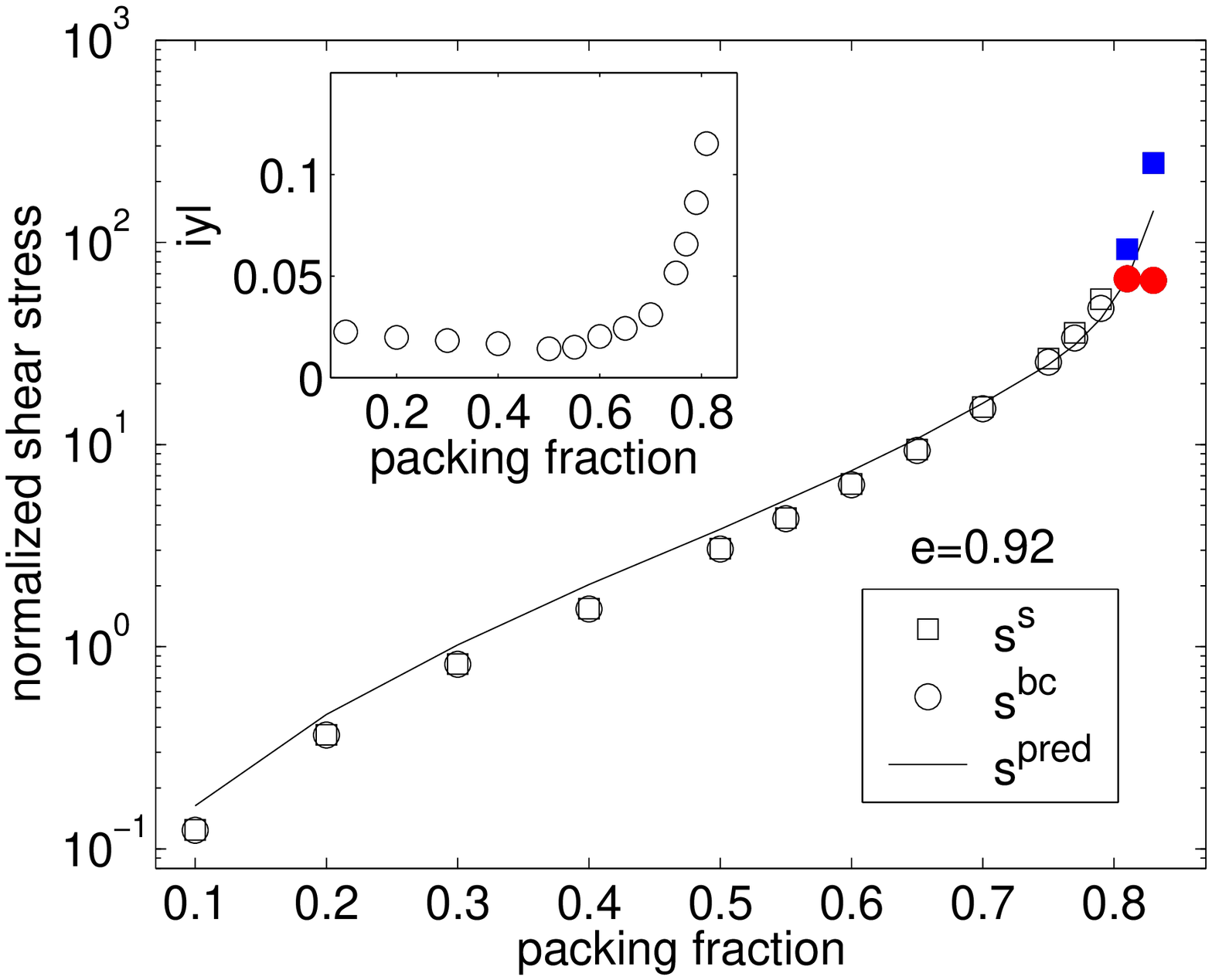}}}
}

\mbox{
\subfigure{\scalebox{.35}{\includegraphics{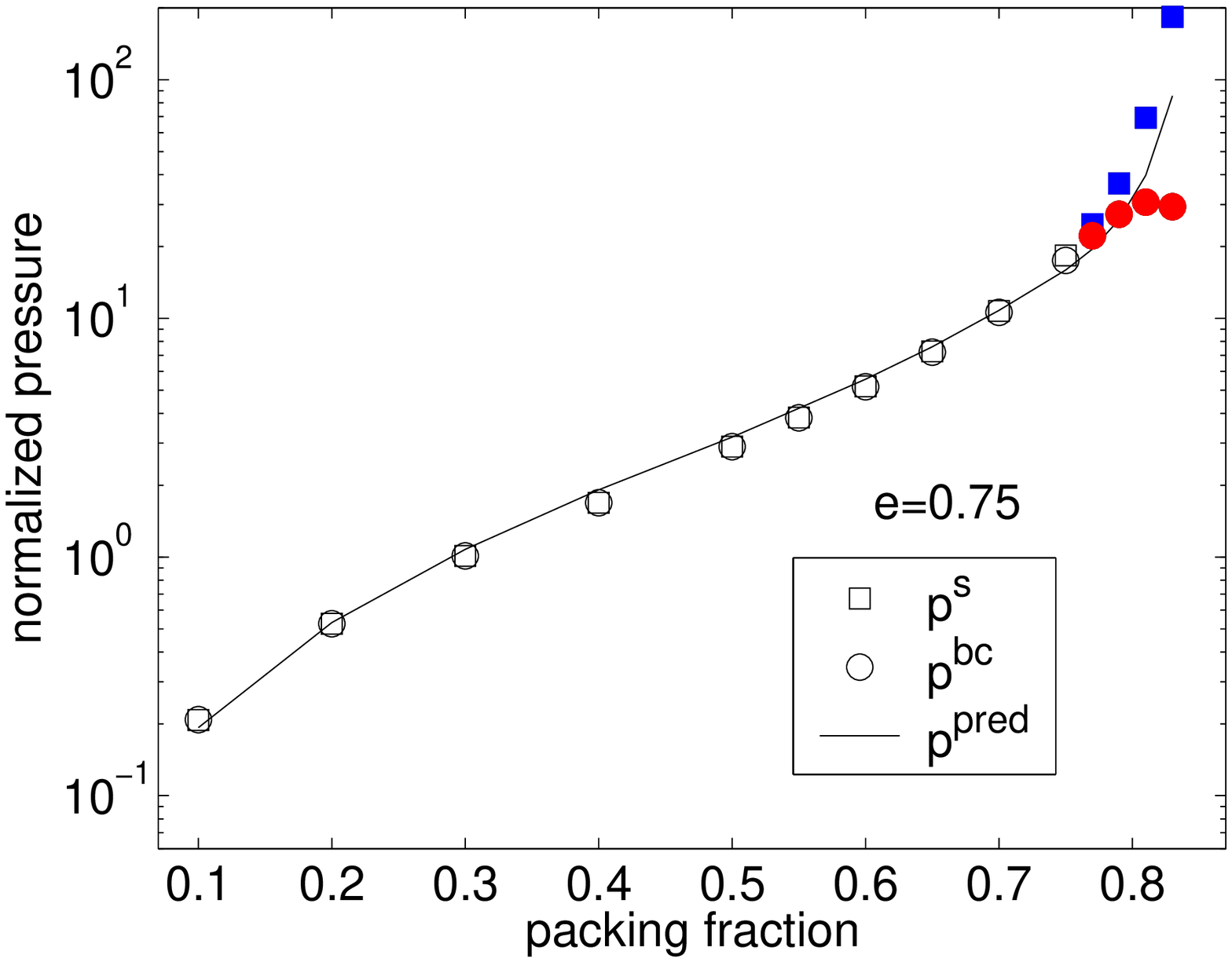}}}
\quad \quad
\psfrag{iyl}{\Huge{$C_v$}}
%\psfrag{ixl}{\Large{packing fraction}}
\subfigure{\scalebox{.35}{\includegraphics{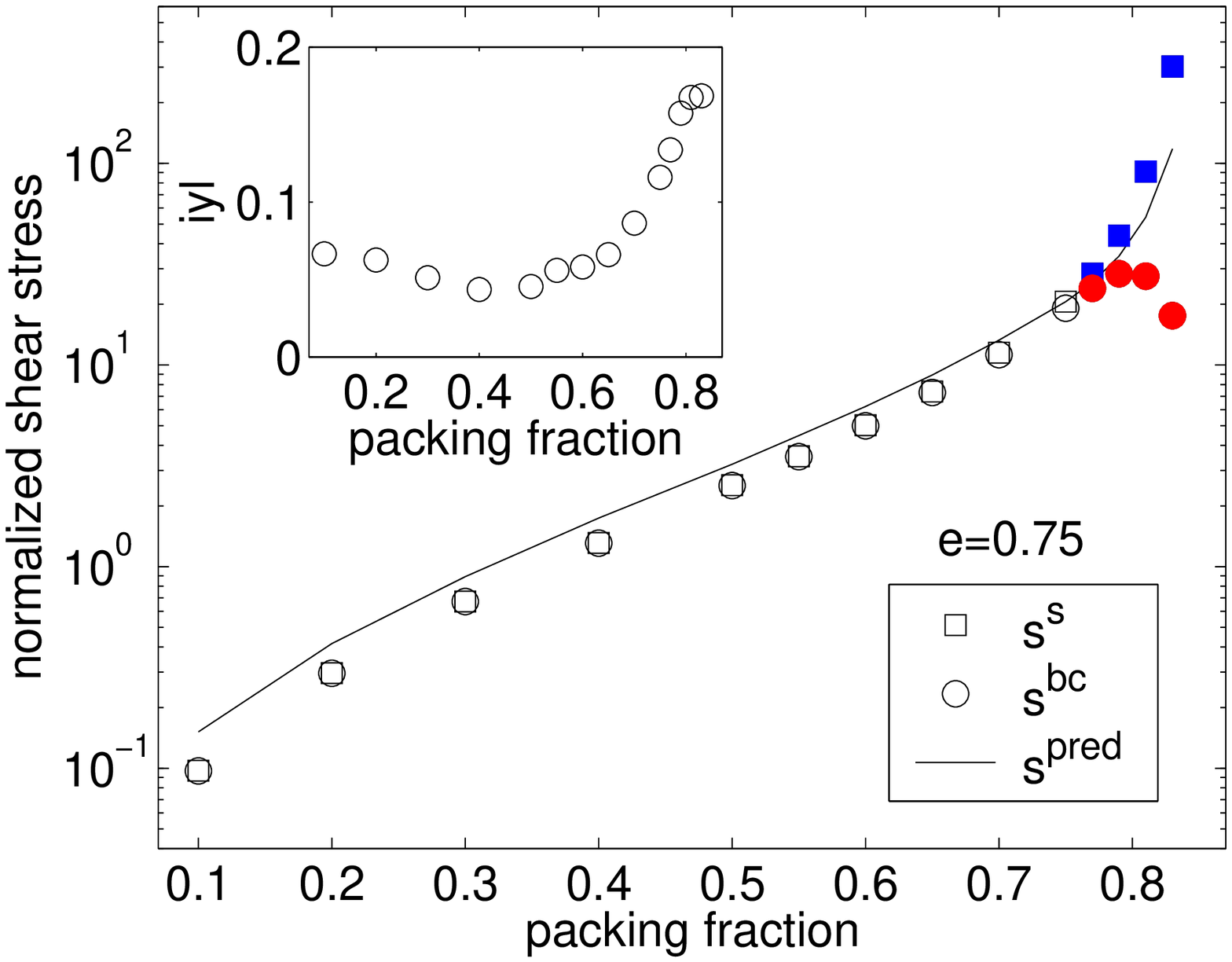}}}
}

\mbox{
\subfigure{\scalebox{.35}{\includegraphics{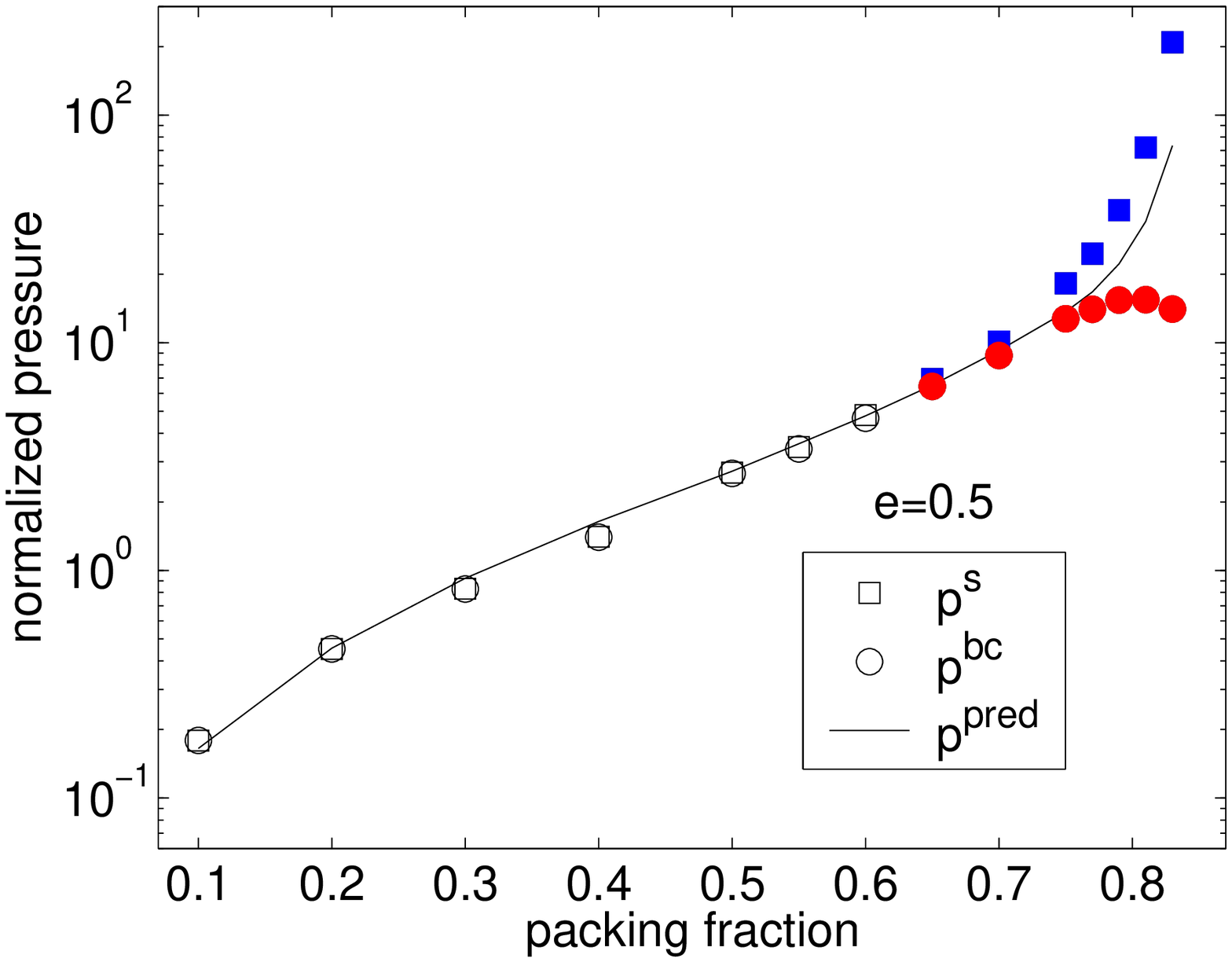}}}
\quad \quad
\psfrag{iyl}{\Huge{$C_v$}}
%\psfrag{ixl}{\Large{packing fraction}}
\subfigure{\scalebox{.35}{\includegraphics{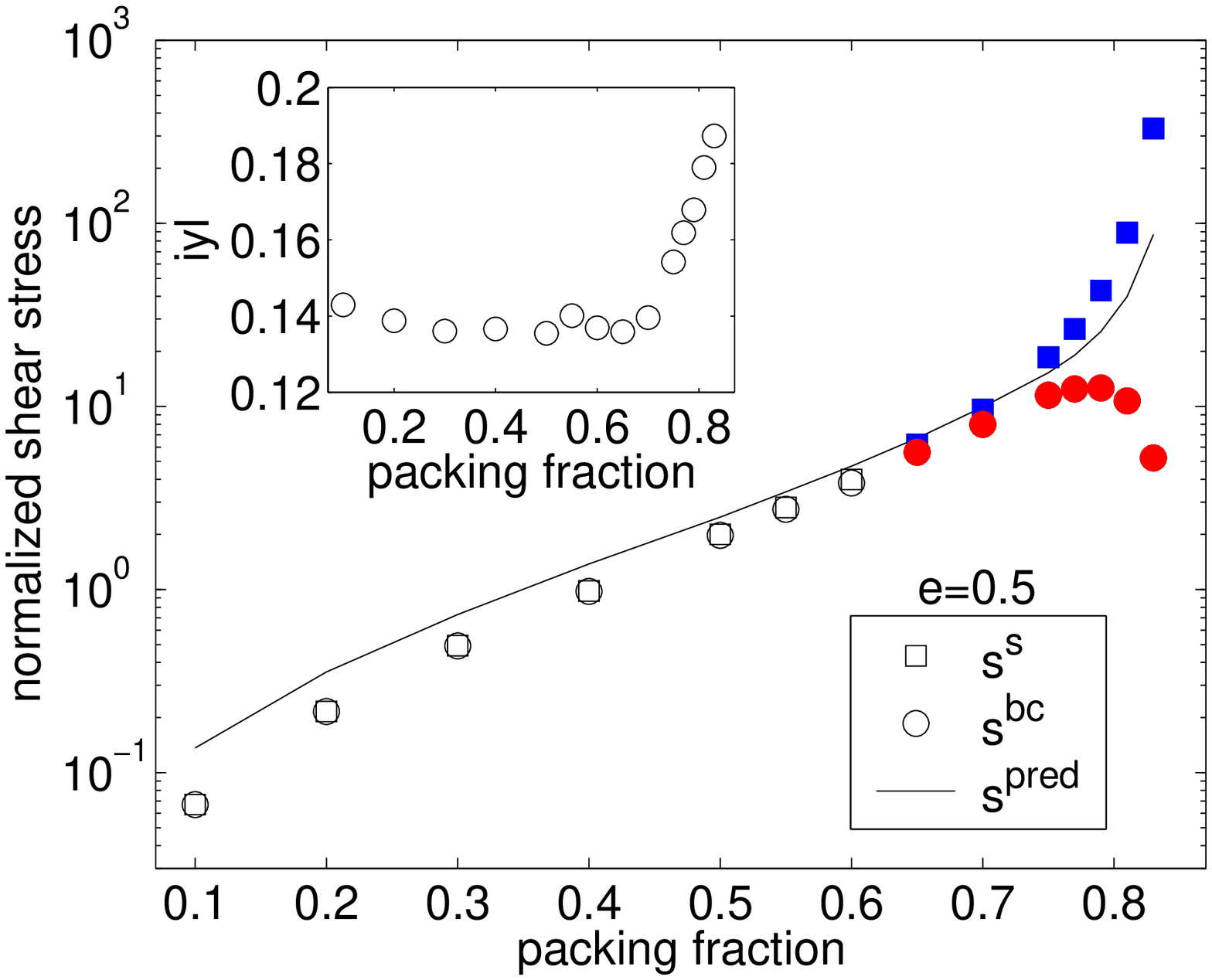}}}
}

%\caption{\label{stresstensorsbige} (Color online).  {\bf Main Figures:}  Normalized values of the pressure (left) and shear stress (right) for large restitution coefficients $e=0.92$, $e=0.75$ and $e=0.5$.  The pressure is normalized by $p_0$ from Equation~(\ref{predpressure}) and the shear stress is normalized by $\dot\gamma \eta_0$ from Equation~(\ref{predshear}) .  The dilute regime is characterized by the range of restitution and packing where the static and collisional values are equal.  Filled data points correspond to values of restitution and packing where $\xi/\xi_\mathrm{el} >1.25$-- this provides a simple quantitative condition for the boundary of the dilute regime.  Kinetic theory is expected to apply in the dilute regime and the prediction for the pressure is accurate for all $e$.  The prediction for the shear stress overestimates the actual value, due to positive velocity correlations.  
%{\bf Insets:} Pre-collisional velocity correlations as a function of packing fraction (defined in Equation~(\ref{precoll})).}
\caption{\label{stresstensorsbige} (Color online).  Main Figures:  Normalized values of the pressure (left) and shear stress (right) for large restitution coefficients $e=0.92$, $e=0.75$ and $e=0.5$.  The pressure is normalized by $p_0$ from Equation~(\ref{predpressure}) and the shear stress is normalized by $\dot\gamma \eta_0$ from Equation~(\ref{predshear}) .  The dilute regime is characterized by the range of restitution and packing where the static and collisional values are equal.  Filled data points correspond to values of restitution and packing where $\xi/\xi_\mathrm{el} >1.25$-- this provides a simple quantitative condition for the boundary of the dilute regime.  Kinetic theory is expected to apply in the dilute regime and the prediction for the pressure is accurate for all $e$.  The prediction for the shear stress overestimates the actual value, due to positive velocity correlations.  
Insets: Pre-collisional velocity correlations as a function of packing fraction (defined in Equation~(\ref{precoll})).}
\end{center}
\end{figure}

\begin{figure}[htbp]
\begin{center}
\mbox{
\subfigure{\scalebox{.35}{\includegraphics{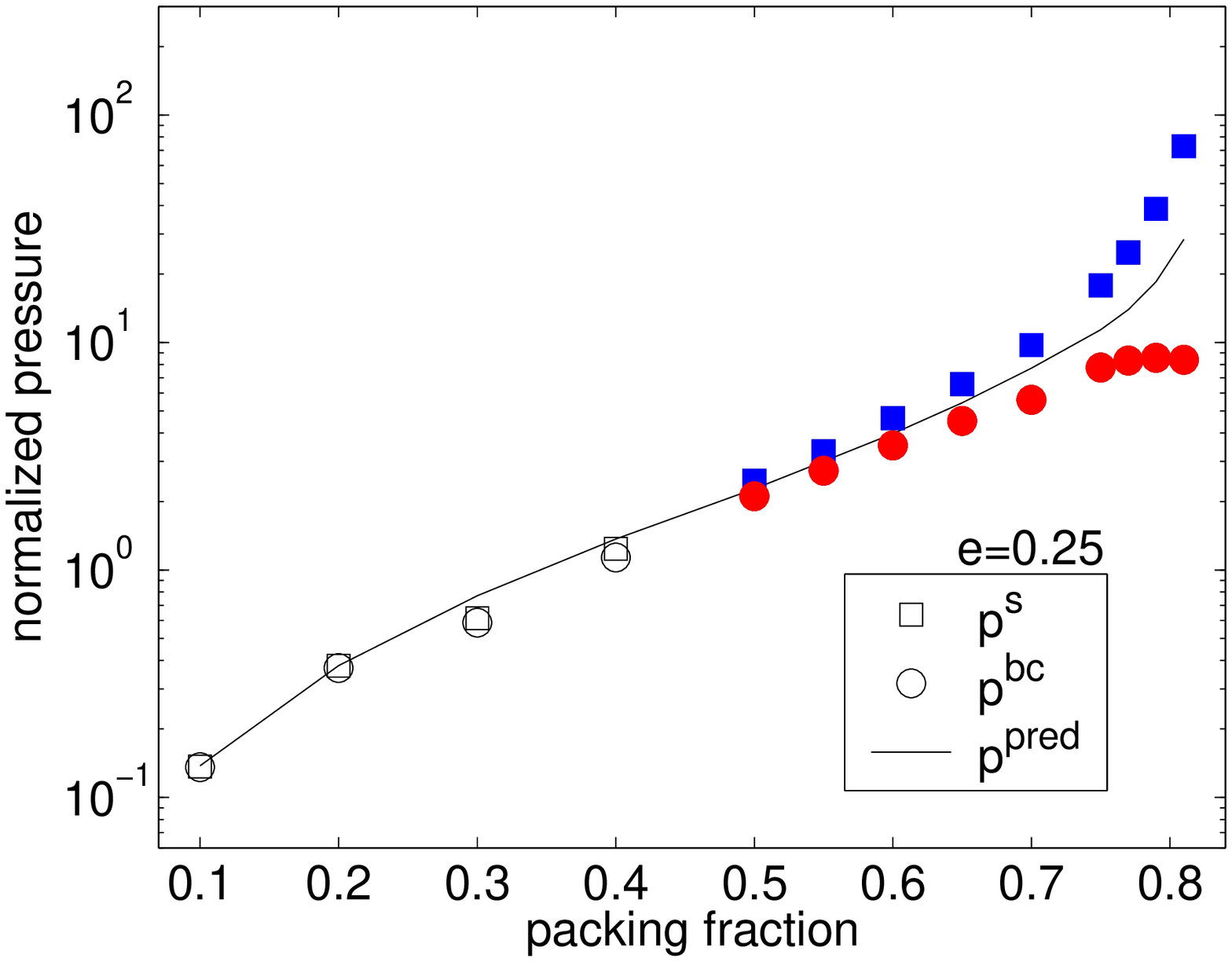}}}
\quad \quad
\psfrag{iyl}{\Huge{$C_v$}}
\subfigure{\scalebox{.35}{\includegraphics{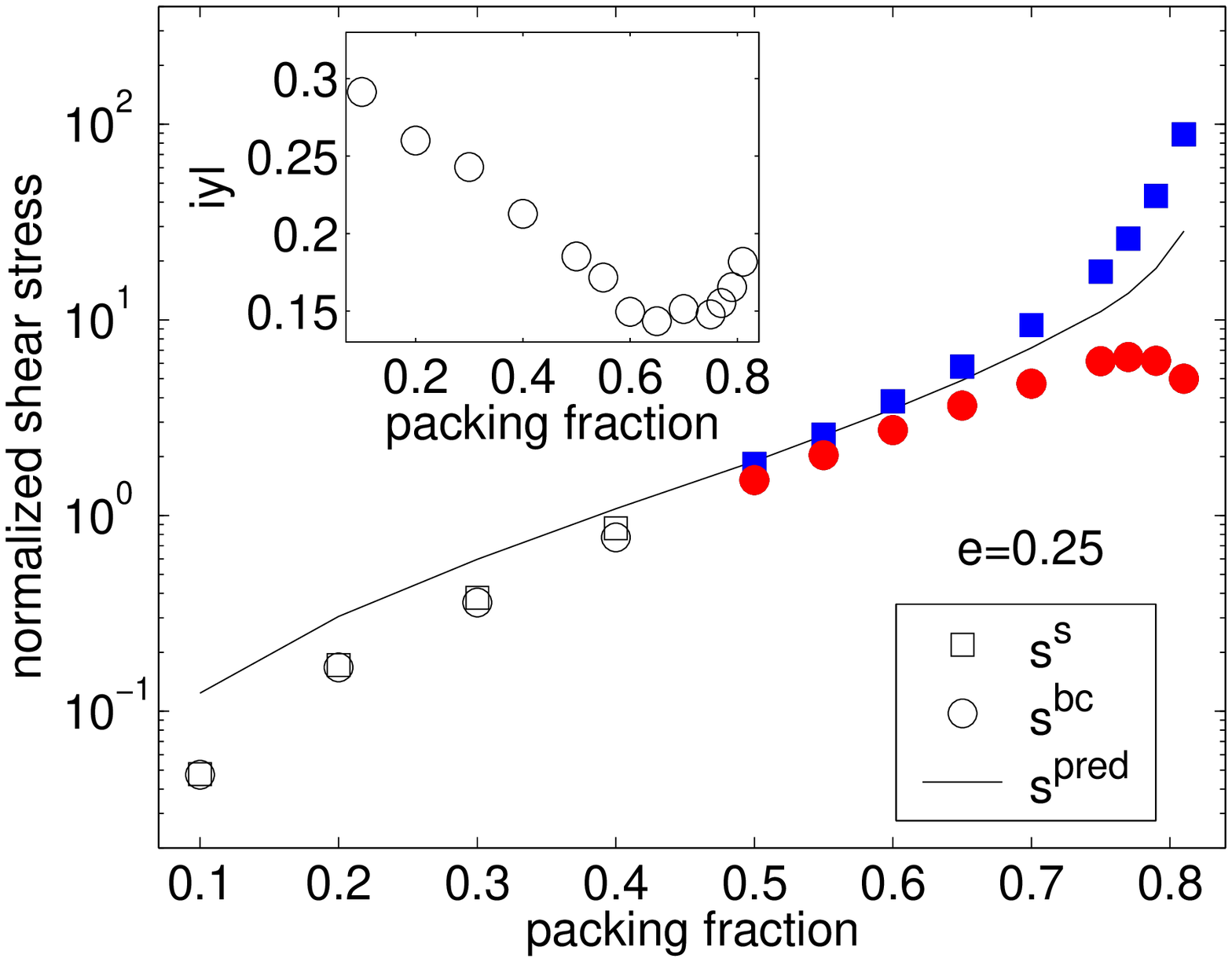}}}
}

\mbox{
\subfigure{\scalebox{.35}{\includegraphics{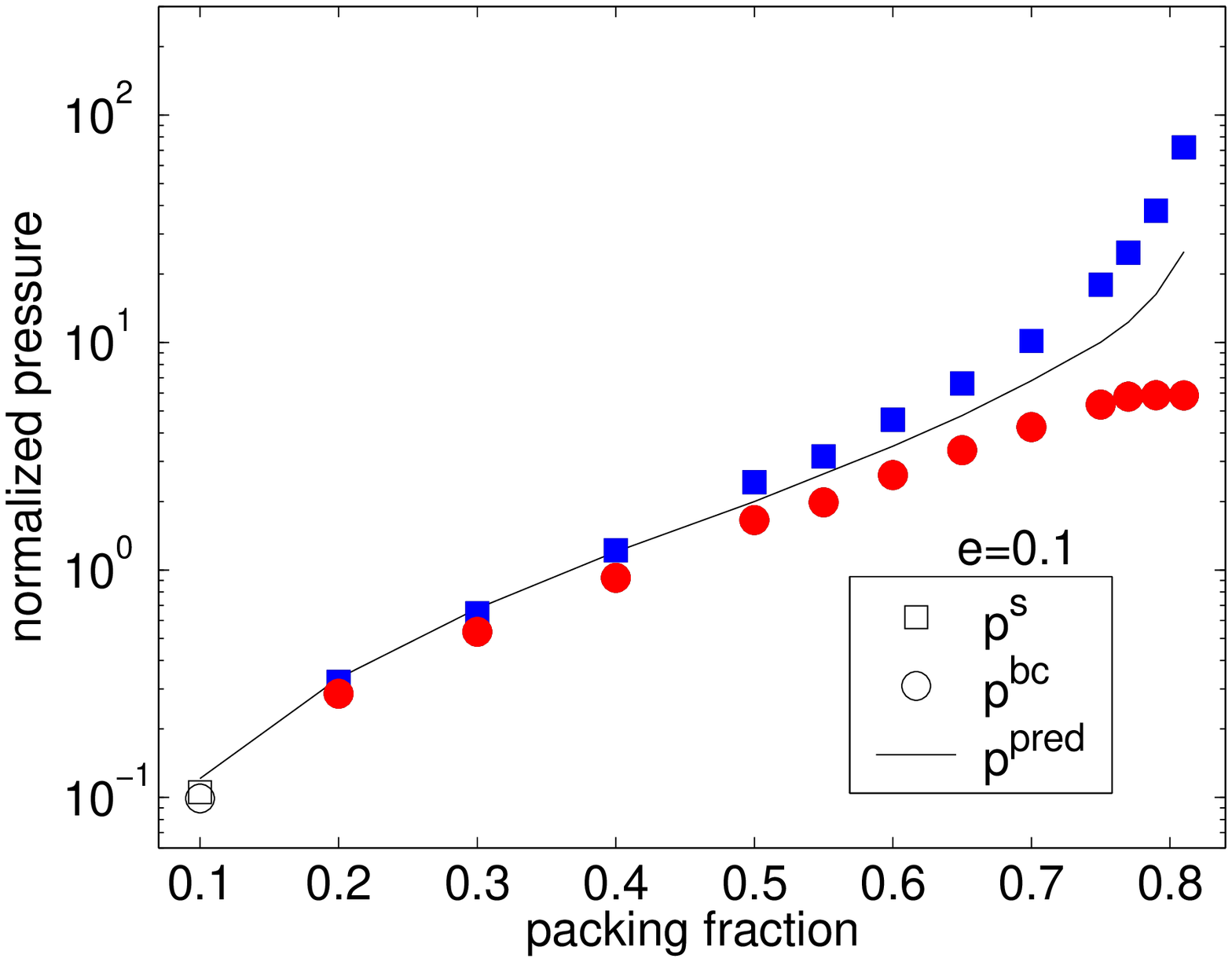}}}
\quad \quad
\psfrag{iyl}{\Huge{$C_v$}}
\subfigure{\scalebox{.35}{\includegraphics{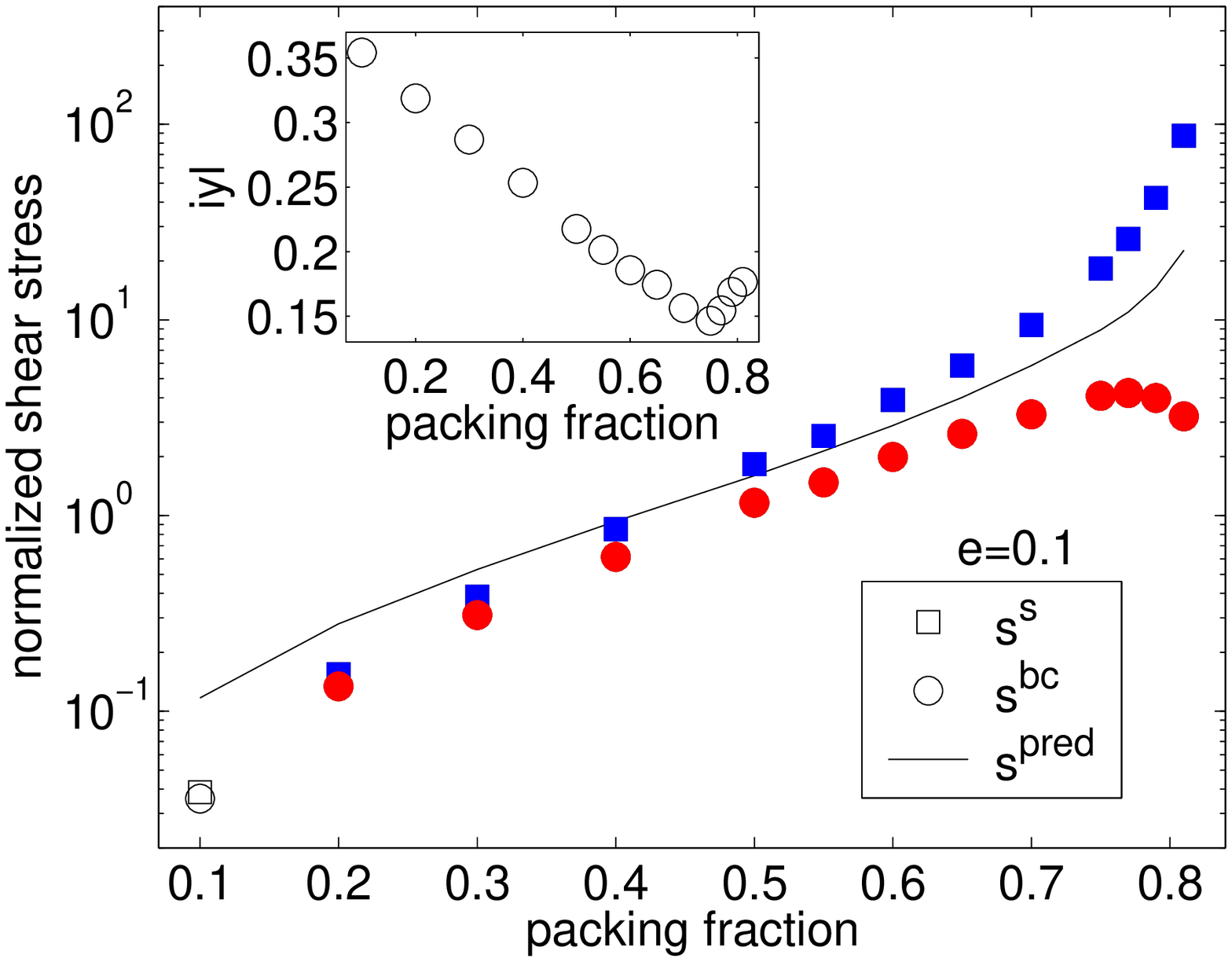}}}
}

\mbox{
\subfigure{\scalebox{.35}{\includegraphics{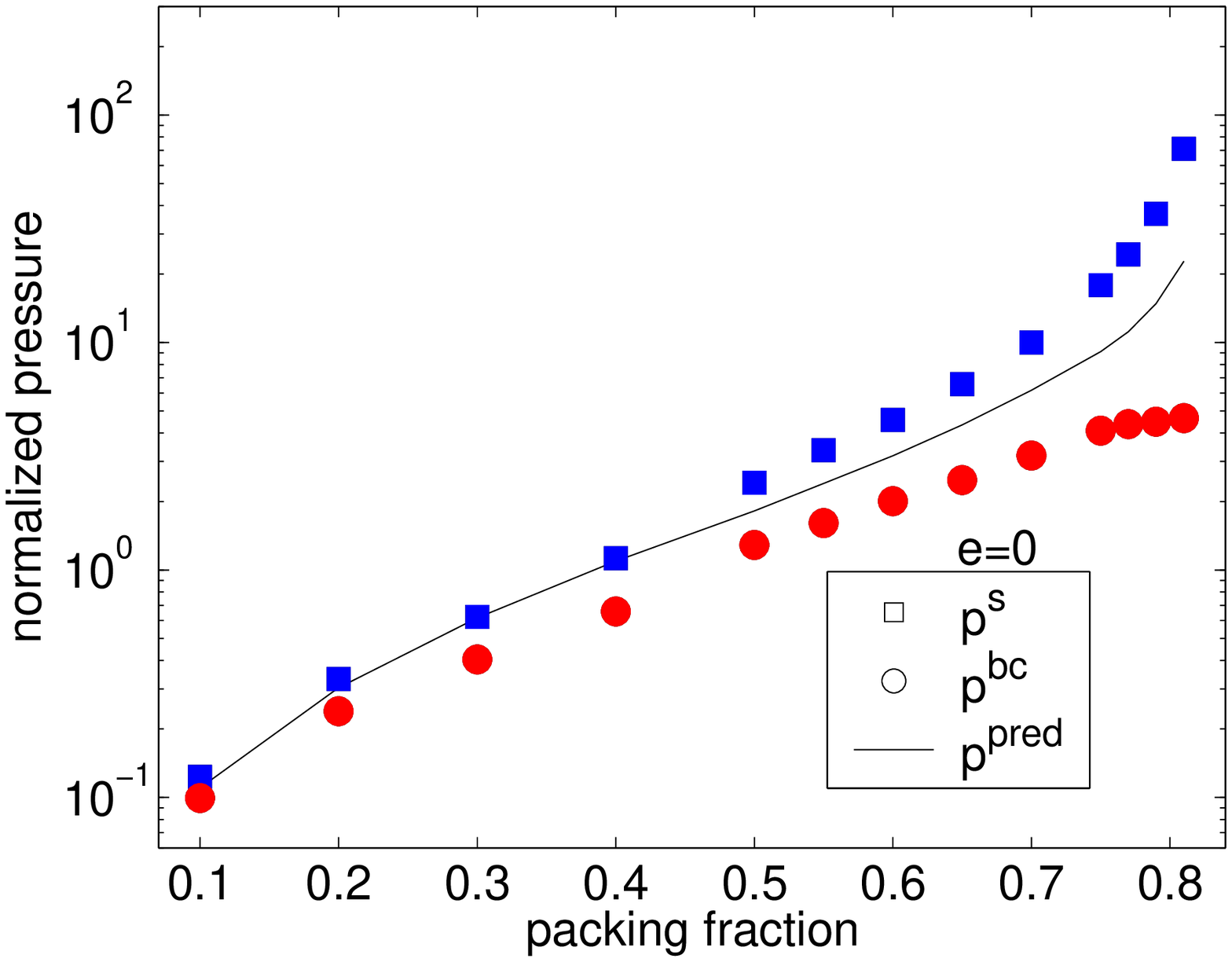}}}
\quad \quad
\psfrag{iyl}{\Huge{$C_v$}}
\subfigure{\scalebox{.35}{\includegraphics{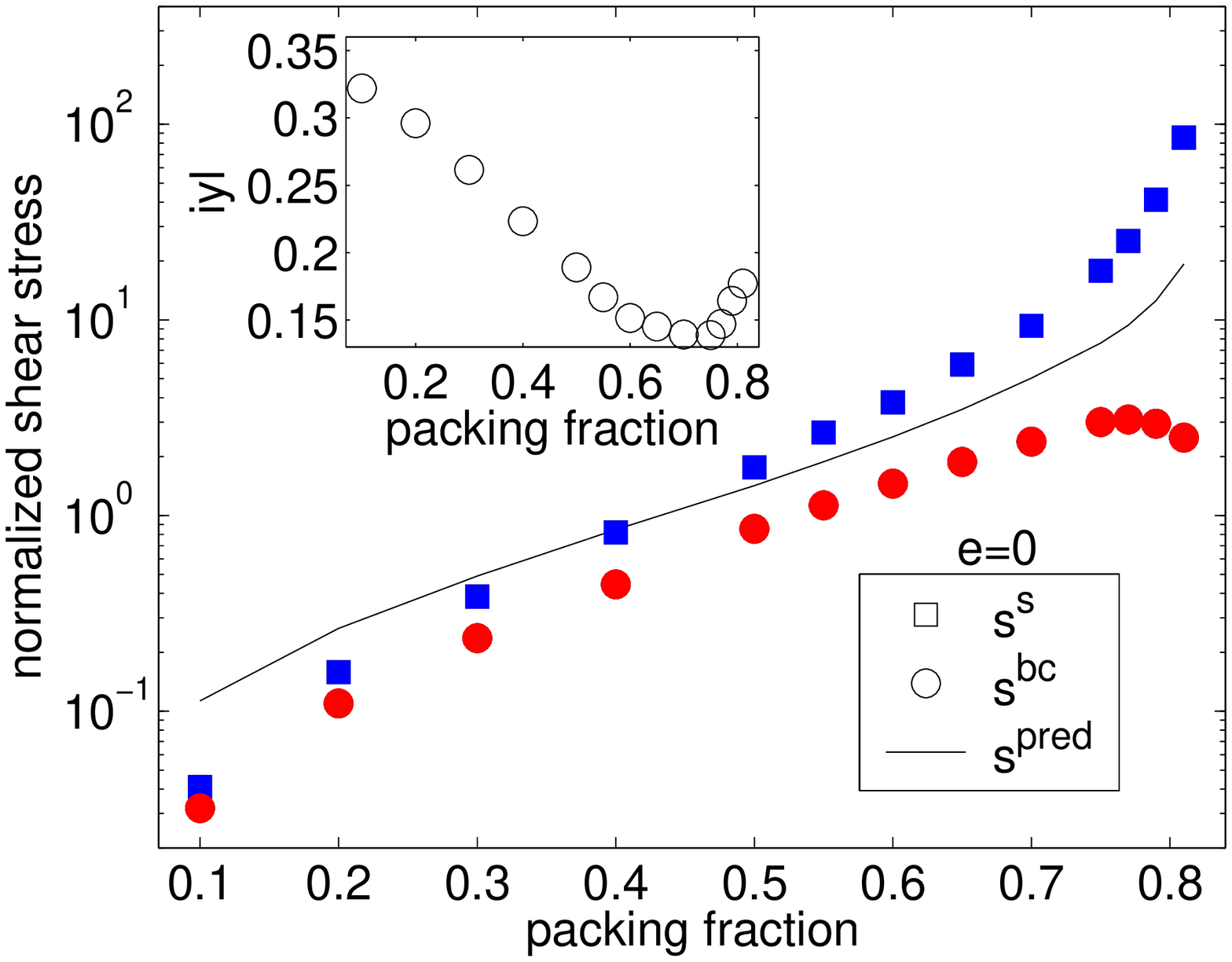}}}
}

\caption{\label{stresstensorssmalle} (Color online).  The same as Figure~(\ref{stresstensorsbige}), for small restitution coefficients of $e=0.25$, $e=0.1$ and $e=0$.}

\end{center}
\end{figure}

In Figures~\ref{stresstensorsbige}~and~\ref{stresstensorssmalle} we plot measurements of the static and collisional stress tensor, for a wide range of restitution coefficients and packing fractions, in terms of the pressure $p$ 
and shear stress $s$, which are related to the stress tensors by   
\begin{eqnarray}
\{ p^{\mathrm{s}} , p^{\mathrm{bc}} \} &=& \{ \frac{1}{2} (\Sigma^{\mathrm{s}}_{11}+\Sigma^{\mathrm{s}}_{22}),\frac{1}{2} (\Sigma^{\mathrm{bc}}_{11}+\Sigma^{\mathrm{bc}}_{22}) \} , \\
\{ s^{\mathrm{s}}, s^{\mathrm{bc}} \} &=& \{ \Sigma^{\mathrm{s}}_{12}, \Sigma^{\mathrm{bc}}_{12} \} =  \{ \Sigma^{\mathrm{s}}_{21}, \Sigma^{\mathrm{bc}}_{21} \}.
\end{eqnarray}

In Figures~\ref{stresstensorsbige}~and~\ref{stresstensorssmalle} both the collisional and static values of the pressure and shear stress are normalized by common factors that are explained later, in the paragraph beneath Equation~(\ref{restfe}).  For now, it is only important to note that there is a regime where the binary collision assumption holds and the normalized stress tensors are equal.  The bounds of this regime depend on the value of both the restitution coefficient and the packing fraction.  The data supports a conclusion that the dilute regime is approached as density is reduced or restitution is increased.

Instead of characterizing the dilute regime in terms of restitution and packing, it is useful to relate it to the length scale $\xi$.  In Figures~\ref{stresstensorsbige}~and~\ref{stresstensorssmalle} we have colored the data points where $\xi/\xi_\mathrm{el}>1.25$.  For all of our data, this simple condition on $\xi$ nicely characterizes the dilute regime-- if $\xi/\xi_\mathrm{el} <1.25$ then the static stress tensor is approximately equal to the collisional stress tensor and the predictions of kinetic theory apply; if $\xi/\xi_\mathrm{el} >1.25$ then interactions between networks of grains become important and kinetic theory does not apply.  The crossover value of $\xi/\xi_\mathrm{el} = 1.25$ is also where the contact force distribution $P(f)$ loses its peak~\cite{gajcompanion1}.  This provides further evidence that $\xi/\xi_\mathrm{el}=1.25$ is a good quantitative bound for the dilute regime.
%Other than the discrepancy between the collisional and static stress tensors in the dense regime, one of the other interesting features of the data in Figures~\ref{stresstensorsbige}~and~\ref{stresstensorssmalle} is the large density behavior of the collisional stress.  For large packing, $s^{\mathrm{bc}}$ actually begins to decrease.  This is expected since, as the density is increased towards the jamming threshold where all velocities become zero, the collisional stress tensor will reduce to zero because it is proportional to the average relative velocity of contacting grains.  
%It is also interesting to note that the collisional shear stress begins to decrease before the collisional pressure.  This is related to the anisotropy of sheared granular materials and the fact that the correlation length depends on the orientation of pairs of grains.

%\clearpage
%\clearpage
%\clearpage

%\clearpage

In the dilute regime where $\xi/\xi_\mathrm{el} <1.25$ and $\Sigma^{\mathrm{s}}_{\alpha \beta} = \Sigma^{\mathrm{bc}}_{\alpha\beta}$, we can test the predictions of hard-sphere kinetic theory for both the pressure and shear stress.  These predictions have been obtained recently~\cite{garzodufty, lutsko1,lutsko2} using the Chapman-Enskog expansion to solve the Enskog equation.  The Enskog equation determines the time dependence of the one-particle probability distribution function (pdf) in terms of collision events between grains.  Collision events consist of binary interactions and the time dependence of the one-particle pdf can therefore be expressed in terms of just the two-particle pdf.  For hard-sphere granular materials, Enskog's equation reads
\begin{equation}
(\partial_t+{\bf v_1} \cdot {\bf \nabla_1}) f^{(1)}({\bf r_1},{\bf v_1},t) = J_E[{\bf r_1},{\bf v_1}],
\label{enskog}
\end{equation}
with $J_E$ given by
\begin{eqnarray}
J_E[{\bf r_1},{\bf v_1}] = \sigma \int d{\bf v_2} \int d\mbox{\boldmath $\hat{\sigma}$} \Theta(\mbox{\boldmath $\hat{\sigma}$} \cdot {\bf g}) (\mbox{\boldmath $\hat{\sigma}$} \cdot {\bf g})   
\label{enskogoperator}
\\
\times [ e^{-2} f^{(2)}({\bf r_1},{\bf r_1}-\mbox{ \boldmath $\sigma$},{\bf v'_1},{\bf v'_2},t) \nonumber
\\
 - f^{(2)}({\bf r_1},{\bf r_1}+\mbox{\boldmath $\sigma$},{\bf v_1},{\bf v_2},t) ]. \nonumber
\end{eqnarray}
Physically, the time dependence of the one-particle pdf $f^{(1)}$ is related to a collisional term $J_E$ that quantifies the probability to gain and lose contributions at a certain velocity ${\bf v_1}$.  The first term in $J_E$ gives the probability that a binary collision between grains results in a grain having velocity ${\bf v_1}$, and the second term gives the probability that a binary collision occurs involving a grain that has velocity ${\bf v_1}$, thereby reducing $f^{(1)}({\bf v_1})$.  The binary collisions occur according to Equation~(\ref{binarycollisions}) and primed velocities represent pre-collisional values. $\Theta$ is the step function and ${\bf g} = {\bf v_1}-{\bf v_2}$.   
For hard sphere granular flows, this form for the Enskog equation can be formally derived, starting with the binary collision assumption and the pseudo-Liouville equation~\cite{vannoijebook, breystat}.

A prediction for the collisional stress tensor is obtained by multiplying Equation~(\ref{enskog}) on each side by $m{\bf v_1}$, where $m$ is the particle mass, and integrating over ${\bf v_1}$.  This yields the transport equation for momentum density~\cite{garzodufty}, which gives the stress tensor
\begin{eqnarray}
\Sigma^{\mathrm{pred}}_{\alpha \beta} &=& \frac{1+e}{4} m \sigma \int d{\bf v_1} \int d{\bf v_2} \int d\mbox{\boldmath $\hat{\sigma}$} \Theta(\mbox{\boldmath $\hat\sigma$} \cdot {\bf g}) (\mbox{\boldmath $\hat\sigma$} \cdot {\bf g})^2 \sigma_\alpha \sigma_\beta \nonumber \\
&\times& \int_0^1 d\lambda f^{(2)}\left[{\bf r}-(1-\lambda) \mbox{\boldmath $\sigma$},{\bf r}+\lambda \mbox{\boldmath $\sigma$},{\bf v_1}, {\bf v_2}, t \right].
\label{formalpredstress}
\end{eqnarray}

In order to determine the stress tensor and solve the Enskog equation, it is necessary to express $f^{(2)}$ in terms of $f^{(1)}$.  Assuming there are no velocity correlations between grains that are about to collide yields
\begin{equation}
f^{[2]}({\bf r_1},{\bf r_2},{\bf v_1},{\bf v_2},t) = \chi({\bf r_1},{\bf r_2}) f({\bf r_1},{\bf v_1},t) f({\bf r_2},{\bf v_2},t)
\label{molecularchaos}
\end{equation}
and reduces the Enskog Equation~(\ref{enskog}) to a non-linear differential equation for the one-particle pdf.  The function $\chi$ is interpreted as the equilibrium correlation function at contact and depends on the local value of the density.  

Once the Enskog equation has been expressed in terms of only the one-particle pdf, it can be solved using the Chapman-Enskog expansion~\cite{ktbook1,ktbook2}, which expands $f^{(1)}$ and $J_E$ in gradients of the mass density, momentum density, and energy density.  This process has been carried out for granular shear flows to first order in the gradients in Refs.~\cite{garzodufty,lutsko1,lutsko2}.  
%When $f^{(1)}$ is determined through this expansion a prediction for the collisional stress tensor, correct through first order in the gradients,  can be made by combining Equations~(\ref{molecularchaos})~and~(\ref{formalpredstress}).  
In two dimensions, this gives predictions~\cite{lutsko2} for the pressure $p^\mathrm{pred}$ and shear stress $s^\mathrm{pred}$
\begin{eqnarray}
\frac{p^{\mathrm{pred}}}{p_0} &=& (1+e) \chi \nu ,
\label{predpressure}
\\
\frac{s^{\mathrm{pred}}}{\dot\gamma \eta_0} &=& \frac{4 \nu}{\pi}  \left( \frac{4}{5-e}+\nu \chi (1+e) f(e) \right),
\label{predshear}
\\
f(e) &=& \frac{3e-1}{5-e}+\left( 1- \frac{(1-e)(1-2e^2)}{81-17e+30e^2(1-e)} \right),
\label{restfe}
\end{eqnarray}
which only depend on the restitution $e$, the packing fraction $\nu$, and the pair correlation function at contact $\chi$.  The normalizing factors are given by $p_0 = n m T / 2$ and $\eta_0 = m/\sigma \sqrt{T/2 \pi}$, where $m$ is the grain mass, $\sigma$ the grain diameter, $T$ the granular temperature, and $n$ the number density.    
We measure all the variables of Equations~(\ref{predpressure}) and (\ref{predshear}) in simulations and test the predictions of hard-sphere kinetic theory without 
fitting parameters.  We use the average value of grain mass and diameter for $m$ and $\sigma$, and determine $\chi$ by tracking the number of collisions that occur per second (which we denote by $\omega$) in equilibrium simulations where $e=1$ and $\dot\gamma=0$~.  Enskog theory relates $\chi$ to $\omega$ through the equation $ \omega = \sqrt{2 \pi \delta T} \chi n \sigma$.  This method for measuring $\chi$ has been used in other recent studies~\cite{lusa}.

We plot the normalized predictions from Equations~(\ref{predpressure}-\ref{restfe}) in Figures~\ref{stresstensorsbige} and \ref{stresstensorssmalle}, where they are compared to data for the stress tensor.  Since hard-sphere kinetic theory assumes binary collisions, the predictions only apply to the dilute regime where $\xi/\xi_\mathrm{el}<1.25$.  These are the open symbols in Figures~\ref{stresstensorsbige} and \ref{stresstensorssmalle}.
We immediately notice that the prediction for the pressure matches the measured pressure in all of the dilute systems we have investigated.  Considering that there are no adjustable parameters, this is a tremendous success for kinetic theory.  

The prediction for the shear stress matches the data in the dilute regime only for large restitution.  As the restitution becomes smaller, the prediction for the shear stress begins to overestimate its measured value.  This overestimation is due to pre-collisional velocity correlations, a mechanism that has been investigated in previous studies~\cite{shattuck}.  Since Equation~(\ref{molecularchaos}) assumes no pre-collisional velocity correlations, if these correlations exist then the average momentum transferred in each collision changes.  If the pre-collisional normal velocities of two grains tend to be aligned (anti-aligned) then the average momentum transferred decreases (increases), causing the kinetic theory prediction to overestimate (underestimate) the measured values.  

In the insets of Figures~\ref{stresstensorsbige}~and~\ref{stresstensorssmalle}, we plot measurements of the pre-collisional velocity correlations $C_v$ defined as 
\begin{equation}
C_v = \langle ( {\bf v'}^i \cdot \mbox{\boldmath $\hat{\sigma}$}^{ij}) ( {\bf v'}^j \cdot \mbox{\boldmath $\hat{\sigma}$}^{ij}) \rangle / T,
\label{precoll}
\end{equation}
which are normalized by the granular temperature.  This definition yields a positive value when pre-collisional grain velocities tend to be aligned, and for all restitution coefficients we observe that the correlations are positive.  In addition, the magnitude of the discrepancy between measured and predicted shear stress is roughly proportional to the size of the velocity correlations.  These observations support the conclusion that the errors in the kinetic theory prediction are due to pre-collisional velocity correlations.  
%It is, however, surprising that the correlations affect the predicted value of the shear stress, but not the predicted value of the pressure.  This seems to be related to an anisotropy in the pre-collisional velocity correlations, although we have not yet pursued this measurement.

To summarize, the stress tensor predicted by hard-sphere kinetic theory matches data from our simulations in the dilute regime where $\xi/\xi_\mathrm{el}<1.25$ and at high restitution coefficients.  As the restitution coefficient is decreased in dilute flows, the prediction for the shear stress begins to fail due to the ``molecular chaos'' assumption of Equation~(\ref{molecularchaos}).  Additionally, as $\xi/\xi_\mathrm{el}$ becomes larger than $1.25$ in the dense regime, the hard-sphere kinetic theory predictions are inaccurate since the binary collision assumption is not valid.  These observations indicate that the fundamental assumptions of molecular chaos and binary interaction do not always apply, and must be addressed.
%These observations demonstrate that the calculations that have been made~\cite{garzodufty, lutsko1, lutsko2} using the present assumptions of hard-sphere kinetic theory are accurate, and it is the nature of the fundamental assumptions which need to be addressed.  

Recent research~\cite{vannoijebook, vannoijepre} has concentrated on refining the molecular chaos assumption to account for velocity correlations in the dilute regime, which have been measured extensively~\cite{shattuck,mesoscopictheory, blairkudrolli, prevost, pouliquenvelcorr}.  However in the dense regime, where networks of interacting grains become important, even an exact inclusion of velocity correlations will not accurately describe the physics since the binary collision assumption does not apply.
%In this case, new theories must be developed that take force correlations and collective motion into account.  
 
\section{Dense Inertial Flows}
\label{forcenetsection}
Dense inertial flows are not quasi-static and cannot be modeled by assuming binary interactions between grains.  They span the range of densities between the dilute and quasi-static regimes, and exhibit properties of both limits.  Like dilute flows, dense inertial flows are characterized by a Bagnold rheology where the stress tensor is proportional to the square of the shear rate~\cite{grestsilbert, dacruz, gaj1}.  However, as in quasi-static flows, the value of the stress tensor also depends on the properties of force chain networks.  In fact, the size of the networks is a parameter that is relevant in all regimes of granular flow.  It varies continuously from the dilute regime, where interactions are binary and the network size is unity, to the quasi-static regime where force chain networks extend over the entire system.  Since the behavior of granular flows can be nicely characterized in terms of the network size, we focus here on the role of force chain networks in determining the value of the stress tensor and construct a force network model of momentum transport in granular materials.  
The basic idea is that forces are transferred elastically through finite sized contact networks and the value of the contact force between any pair of grains depends on both the relative velocity of the contacting pair (a ``collisional'' contribution) and the values of the other contact forces in the network, even those a long distance away (an ``elastic'' contribution).  This leads to a prediction for the stress tensor that holds over the entire inertial regime and incorporates the non-collisional stress that arises when force networks have formed.  

\subsection{The Force Network Model}
A central feature of dense granular materials is that forces can be transferred over distances much larger than the grain size.  This is especially evident in static packings of grains, but is also important when considering dense inertial flows.  The only necessary requirement for force propagation is the existence of connected networks of interacting grains.  When this requirement is met, elastic waves propagate through the networks at a speed set by the values of the moduli and there is an elastic component of the stress tensor.  This elastic contribution must be added to the collisional part of the stress in Equation~(\ref{bcstress}).  Therefore the full static stress tensor is comprised of two terms:  one describing the elastic response and one describing the collisional response.  

From a microscopic viewpoint, the value of the stress tensor is determined by contact forces between grains, as in Equation~(\ref{sstress}).  Therefore, in the presence of force networks, these contact forces must also be comprised of collisional and elastic terms.  The collisional term is given by Equation~(\ref{binaryforce}) and represents the local force due to collisions between pairs of grains.  It depends only on properties of the two contacting grains.  The elastic term is a result of elastic deformations in the contact network.  It is a non-local force that arises from the network applying an effective pressure on every pair of contacting grains.   

In dilute flows only the collisional part of contact forces is non-zero and the stress tensor is well described by kinetic theory.  In the quasi-static regime the elastic part of the forces is much larger than the collisional part, and the latter is usually disregarded.  In dense inertial flows, both terms contribute.  The force network model extends hard-sphere kinetic theory by explicitly calculating the effects of elastic waves in force networks.  This leads to a prediction for the elastic portion of contact forces, which result from forces propagating through force networks at the elastic wave speed.  Inertial flows correspond to the limit where forces propagate throughout the entire network before it is destroyed.  This is equivalent to the limit where the elastic wave speed is infinite, or the grains are perfectly rigid.  

\begin{figure}[htbp]
\begin{center}
\subfigure{\scalebox{0.37}{\includegraphics{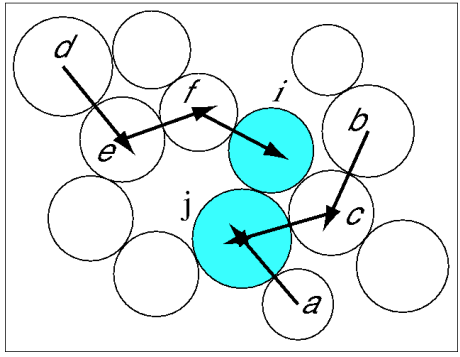}}}
\caption{\label{networkpic} A network of contacting grains.  The contact force between the shaded grains $i$ and $j$ is determined by the local collisional force $F^{ij}_{\mathrm{bc}}$, plus elastic contributions from forces propagating through the network.  It is convenient to organize these non-local effects into contributions from different path-lengths.  On the top left is a path of size three where the local contact force between grains $d$ and $e$ is transferred through the network to grain $i$.  On the top right is a path of size two, and on the bottom right is a path of size one.    
}
\end{center}
\end{figure}

Mathematically, the contact force $F^{ij}$ between grains $i$ and $j$ is equal to the sum of a collisional term, plus elastic effects from the network:
\begin{equation}
F^{ij} = F^{ij}_{\mathrm{bc}} + \sum_{\ell = 1}^{\ell_\mathrm{max}} \mathcal{F}^{ij}_\ell.
\label{basicforceprop}
\end{equation}  
In this equation the first term is the collisional force, defined in Equation~(\ref{binaryforce}), and the second term arises from forces that propagate through paths in the force chain network, as illustrated in Figure~\ref{networkpic}.  We split this term into contributions $\mathcal{F}^{ij}_\ell$ that represent added forces from different path-lengths $\ell$.  The total elastic force created by the network is therefore equal to the sum of the contributions $\mathcal{F}^{ij}_\ell$ over all possible path-lengths $\ell<\ell_\mathrm{max}$ in the force network.

Figure~\ref{networkpic} illustrates force propagation and defines the notion of path-length.  For example, because grain $d$ is in contact with grain $e$, this increases the force between grains $e$ and $f$, which increases the force between grains $f$ and $i$, which has the net effect of increasing the contact force between grains $i$ and $j$.  We denote this as a path of length three ($\ell=3$) since the local force $F^{de}_{\mathrm{bc}}$ must propagate through three links to influence the contact between $i$ and $j$.  

The net effect of forces propagating through different path-lengths can be calculated explicitly in the limit of large elastic wave speeds.  We begin by considering a path of length $\ell=1$, as illustrated in Figure~\ref{networkpic}:
because grain $a$ is in contact with grain $j$, the local contact force $F^{aj}_{\mathrm{bc}}$ increases the value of $F^{ij}$ by an amount equal to $F^{aj}_{\mathrm{bc}}$ multiplied by the cosine of the angle between the unit vectors connecting contacts $\{a,j\}$ and $\{i,j\}$.
If we assume that grain $i$ has $z_i$ contacts labeled by $m$ and grain $j$ has $z_j$ contacts labeled by $n$, then the effect of all paths of length one is to increase $F^{ij}$ in Equation~(\ref{basicforceprop}) by an amount 
\begin{equation} 
 \mathcal{F}^{ij}_1 =  \sum_{m=1; m\neq j}^{z_i}{ \mbox{\boldmath $\hat{\sigma}$}^{mi} \cdot \mbox{\boldmath $\hat{\sigma}$}^{ij} F^{mi}_{\mathrm{bc}}} + \sum_{n=1; n\neq i}^{z_j}{ \mbox{\boldmath $\hat{\sigma}$}^{nj} \cdot \mbox{\boldmath $\hat{\sigma}$}^{ij} F^{nj}_{\mathrm{bc}}} ,
\label{pathone}
\end{equation}
where $\mbox{\boldmath $\hat\sigma$}^{ab}$ is the unit vector connecting the center of grains $a$ and $b$.  This expression includes all of the effects from paths of length $\ell=1$ on each of the contacting grains $i$ and $j$.  

In an analogous manner, the path from grain $b$ to $c$ to $j$ in Figure~\ref{networkpic} comprises a path of length two ($\ell=2$), which increases the value of $F^{ij}$ due to the local force between $b$ and $c$.  Note that we ignore the fact that the local force between $c$ and $j$ also increases $F^{ij}$, since this contribution was already included in Equation~(\ref{pathone}).  The total additional force between grains $i$ and $j$ arising from paths of length two is given by
\begin{eqnarray}
 \mathcal{F}^{ij}_2 &=& \sum_{m=1; m\neq j}^{z_i}{ \mbox{\boldmath $\hat{\sigma}$}^{mi} \cdot \mbox{\boldmath $\hat{\sigma}$}^{ij} \sum_{p=1; p \neq m}^{z_m}{  \mbox{\boldmath $\hat{\sigma}$}^{pm} \cdot \mbox{\boldmath $\hat{\sigma}$}^{mi} F^{pm}_{\mathrm{bc}}}} \nonumber \\
&+& \sum_{n=1; n\neq i}^{z_j}{ \mbox{\boldmath $\hat{\sigma}$}^{nj} \cdot \mbox{\boldmath $\hat{\sigma}$}^{ij} \sum_{q=1; q \neq n}^{z_n}{ \mbox{\boldmath $\hat{\sigma}$}^{qn} \cdot \mbox{\boldmath $\hat{\sigma}$}^{nj} F^{qn}_{\mathrm{bc}}}} ,
\label{pathtwo}
\end{eqnarray}
where once again grain $i$ has $z_i$ contacts labeled by $m$ and grain $j$ has $z_j$ contacts labeled by $n$.  To calculate the effect of paths of length two, we also take
into account the $z_m$ contacts of grain $m$, labeled by $p$, and the $z_n$ contacts of grain $n$, labeled by $q$. 

The contributions to $F^{ij}$ from arbitrary path-lengths $\ell$ can be determined by continuing the above arguments.  They depend on the coordination number $z$ and are also sensitive to the geometric arrangement of force networks.  
One important constraint arises as $z$ becomes large, which is a result of energy conservation.  
If we consider the total energy $T^{ij}$ that an arbitrary pair of contacting grains $i$ and $j$ contribute to the network, this must always be less than or equal to the total kinetic energy initially stored in the contact (via the kinetic energies of grains $i$ and $j$).  This is because, while energy is conserved in this process, some is transferred to the elastic energy of the network and some remains as kinetic energy of grains $i$ and $j$.  The average collisional force $\langle F^{ij}_\mathrm{bc} \rangle$ is proportional to the square of the relative velocities~\cite{garzodufty} and thereby proportional to the kinetic energy of the contact.  Therefore, on average, the {\em magnitude} of the collisional force transferred to nearest neighbors must not be greater than $\langle F^{ij}_\mathrm{bc} \rangle$.            
%When considering this constraint it is important to make the distinction between the collisional force that a contact transmits to the network and the collisional force that a contact supplies to the network.  
%Contacts can transmit forces that originate in different contacts, and can also supply the collisional force that originates in the contact itself.  
%The force transmitted through a contact originates in the collisional forces of other contacts, and can be non-zero even in the limit that the relative velocity between the contacting pair $\{i,j\}$ becomes zero and $F_\mathrm{bc}^{ij}=0$.  Transmission occurs according to Equations~(\ref{pathone}) and its larger $\ell$ analogs.  
%A contact can also supply its collisional force to nearest neighbors, which will only occur if $F_\mathrm{bc}^{ij}>0$.  
%In this case, the contact supplies a certain amount of kinetic energy to the network.  The kinetic energy is proportional to the square of the particle velocities, which is proportional to $F_\mathrm{bc}^{ij}$.  If we consider the total energy $T^{ij}$ that an arbitrary pair of contacting grains $i$ and $j$ supplies to the network, this must always be less than or equal to the initial energy, which is proportional to the collisional force $F^{ij}_\mathrm{bc}$.  %This is because, in the hard-sphere limit, the magnitude of the collisional force $F_\mathrm{bc}^{ij}$ is proportional to the energy stored in the contact.  
This leads to the constraint equation
\begin{eqnarray}
T^{ij} &\equiv& \Big\langle F^{ij}_\mathrm{bc} \big( \sum_{m=1; m\neq j}^{z_i}{ \mbox{\boldmath $\hat{\sigma}$}^{mi} \cdot \mbox{\boldmath $\hat{\sigma}$}^{ij}} + \sum_{n=1; n\neq i}^{z_j}{ \mbox{\boldmath $\hat{\sigma}$}^{nj} \cdot \mbox{\boldmath $\hat{\sigma}$}^{ij}} \big) \Big\rangle
\nonumber
\\
& \leq & \langle F^{ij}_\mathrm{bc} \rangle , 
\label{constraintone}
\end{eqnarray}
which is a global constraint.  It ensures that the total energy supplied to the network never exceeds the initial kinetic energy as the elastic waves move from first nearest neighbors, to second nearest neighbors, and so on.   

Equations~(\ref{pathone}-\ref{constraintone}) model the physical origin of elastic forces that exist in dense granular materials and allow for a complete determination of $\mathcal{F}_\ell^{ij}$.  The numerical value of the elastic force between a pair of contacting grains is calculated by summing these contributions over all possible path-lengths $\ell$, as in Equation~(\ref{basicforceprop}).  

The maximum possible path length $\ell_\mathrm{max}$ is constrained by the size of the force networks.  A straight chain of grains that spans the network has $\ell^* = \xi/\xi_\mathrm{el}-1$.  While there are also network spanning chains with $\ell > \ell^*$, their contributions to elastic forces are diminished since the magnitude of the collisional force at the end of the chain is reduced by the product of the cosine of the angles in the chain.  We will therefore set $\ell_\mathrm{max} = \xi/\xi_\mathrm{el}-1$ and only consider the network spanning chains that give the largest contribution.  This amounts to completely ignoring the effects of path lengths with $\ell > \ell_\mathrm{max}$ which increase the elastic force on the contact $\{i,j\}$ by an amount $\Delta = \sum_{\ell=\ell_\mathrm{max}+1}^{\infty} \mathcal{F}^{ij}_\ell$. We do not expect this approximation to produce a large error in the total elastic force on $\{i,j\}$ since $\Delta < \mathcal{F}^{ij}_{\ell_\mathrm{max}}$ and, in general, $\mathcal{F}^{ij}_{\ell+1} \ll \mathcal{F}^{ij}_\ell$ for $\ell < \ell_\mathrm{max}$.

Given the above analysis, it is possible to completely determine the stress tensor based on properties of the force network and the collisional stresses.  This is carried out in Subsection~\ref{calcsection} and the predictions are tested in Subsection~\ref{testsection}.  

\subsection{Calculating the stress tensor}
\label{calcsection}
We calculate the stress tensor by rewriting the equations of the force network model in terms of integrals instead of sums.  Here we carry out this substitution for a two-dimensional system, although it can be generalized to higher dimensions.  If we consider the average force between two grains that contact at an angle $\theta$, denoted $F(\theta)$, then Equation~(\ref{basicforceprop}) can be generalized to read     
\begin{equation}
\label{forceproptheta}
F(\theta) = F_{\mathrm{bc}}(\theta) + \sum_{\ell = 1}^{\xi/\xi_\mathrm{el}-1} \mathcal{F}_\ell(\theta).
\end{equation}
This equates the average force between grains contacting at angle $\theta$ to the average collisional force at that angle, plus elastic effects from the network.  In what follows we will measure $\theta$ with respect to the axis of shear, so that $\theta=0$ corresponds to the axis where there is no gradient in grain velocities.  

Generalizing Equations~(\ref{pathone}) and (\ref{pathtwo}) to arbitrary path-length $\ell$ gives
\begin{eqnarray}
\mathcal{F}_\ell(\theta_0) &=& \prod_{i=1}^{\ell} \int_{\theta_{i-1}-2 \pi/3}^{\theta_{i-1}+2 \pi/3} d\theta_{i} \cos(\theta_i - \theta_{i-1}) C(\theta_i) F_\mathrm{bc}(\theta_\ell)  \nonumber \\
&\times& \Big( \sum_{z(\theta_i)} p\big( z(\theta_i)\big) \big(z(\theta_i)-1\big) P_{z(\theta_i)}(\theta_i-\theta_{i-1}) \Big).
\label{appyoanael}
\end{eqnarray}
In this equation, the sums from the Equations~(\ref{pathone}) and (\ref{pathtwo}) have been replaced by integrals over $\theta_i$, which is the angular orientation of each link in the chain.  Each integral contains a cosine that replaces the dot product.  Note that the bounds of the integrals are arranged such that the grain at link $i$ is not permitted to overlap the grain at link $i-1$.  The function $F_\mathrm{bc}(\theta_\ell)$ provides the collisional force at the end of the path.
In order to properly characterize the probability to have a contact at $\theta_i$, we also introduce the functions $z(\theta)$, $C(\theta)$ and $P_{z(\theta)}$.  $C(\theta)$ gives the probability to have a single contact at angle $\theta$~\cite{thornton,arthur,rothenburg,anisradjai,kruyt}.  In the case that there are two (or more) contacts on a single grain, which is necessary to form a chain, this probability must be modified~\cite{radjai3}.  The function $C(\theta_i) P_{z(\theta_i)}(\theta_i-\theta_{i-1})$ gives the conditional probability to have a contact at $\theta_i$, provided that there already is a contact at $\theta_{i-1}$ and there are a total of $z(\theta_i)$ contacts on the grain at $\theta_i$.  The function $p\big(z(\theta_i)\big)$ is the probability to have a grain with $z(\theta_i)$ contacts, and the term in the parentheses captures the average effect of all possible combinations of contact numbers and angles on each grain in the force chain.

It is useful to simplify Equation~(\ref{appyoanael}) by defining the function $P_C(\theta_i-\theta_{i-1})$, where $C(\theta_i) P_C(\theta_i-\theta_{i-1})$ gives the probability to have a contact at $\theta_i$, given that a contact already exists at $\theta_{i-1}$, averaged over all possible fluctuations in $z(\theta_i)$.  If we define
\begin{equation}
(z-1) P_C(\theta_i-\theta_{i-1}) \equiv  \sum_{z(\theta_i)} p\big(z(\theta_i)\big) \big(z(\theta_i)-1\big) P_{z(\theta_i)}(\theta_i-\theta_{i-1}),
\label{definepc}
\end{equation}
where $z$ is the average coordination number in the entire network, then Equation~(\ref{appyoanael}) simplifies to 
\begin{equation}
\mathcal{F}_\ell(\theta_0) = \prod_{i=1}^{\ell} (z-1) \int_{\theta_{i-1}-2 \pi/3}^{\theta_{i-1}+2 \pi/3} d\theta_{i} \cos(\theta_i - \theta_{i-1}) P_C(\theta_i - \theta_{i-1}) C(\theta_i) F_\mathrm{bc}(\theta_\ell).
\label{appyo}
\end{equation}
Equation~(\ref{appyo}) generalizes the sums in Equations~(\ref{pathone}) and (\ref{pathtwo}) to arbitrary path length $\ell$.  It does so by averaging over one-particle distribution functions.  These come in the form of the average collisional force $F_\mathrm{bc}(\theta)$, the average contact probability on a single grain $C(\theta_i) P_C(\theta_i-\theta_{i-1})$, and the average coordination number $z$.  Naturally, there must not be correlations between these variables in order for the predictions to apply.  The definition of $P_C$ ensures that there are no correlations between the contact probabilities and the coordination number, but it is not guaranteed that the collisional forces are not correlated with the network structure.  In fact, when closed loops form in the force networks, the collisional forces and network structure do become correlated.  Therefore, Equation~(\ref{appyo}) only applies as a mean-field approximation that ignores the correlations induced from loops in the force networks.  This is a valid approximation since the effect of a collisional force that propagates around a loop is reduced by the product of the cosine of the angles around the loop, which is quite small compared with direct propagation.

To solve the integrals in Equation~(\ref{appyo}), we change integration variables to $x_i=\theta_i-\theta_{i-1}$.  This results in the expression
\begin{equation}
\label{appyo1}
\mathcal{F}_\ell(\theta) = \prod_{i=1}^\ell (z-1) \int_{-2 \pi/3}^{2 \pi/3} dx_{i} \cos(x_i) P(x_i) C\bigg[\theta + \sum_{j=1}^{i} x_j\bigg] F_\mathrm{bc}\bigg[\theta + \sum_{j=1}^\ell x_j\bigg],
\end{equation}
which incorporates the average effect of all forces that propagate through paths of length $\ell$ on contacts with orientation $\theta$.  

In addition to force propagation based on the cosine of the angle between subsequent contacts, we must also incorporate the global constraint from Equation~(\ref{constraintone}).  This can be generalized to
\begin{equation}
\label{constrainttwo}
T(\theta) = \langle F_\mathrm{bc}(\theta) \rangle (z-1) \int_{-2 \pi/3}^{2 \pi/3} dx P_C(x) \cos(x) C(\theta+x) \leq \langle F_\mathrm{bc}(\theta) \rangle,
\end{equation}
and restricts the total energy transferred through the network.

\begin{figure}
\begin{center}
\mbox{
\scalebox{0.47}{\includegraphics{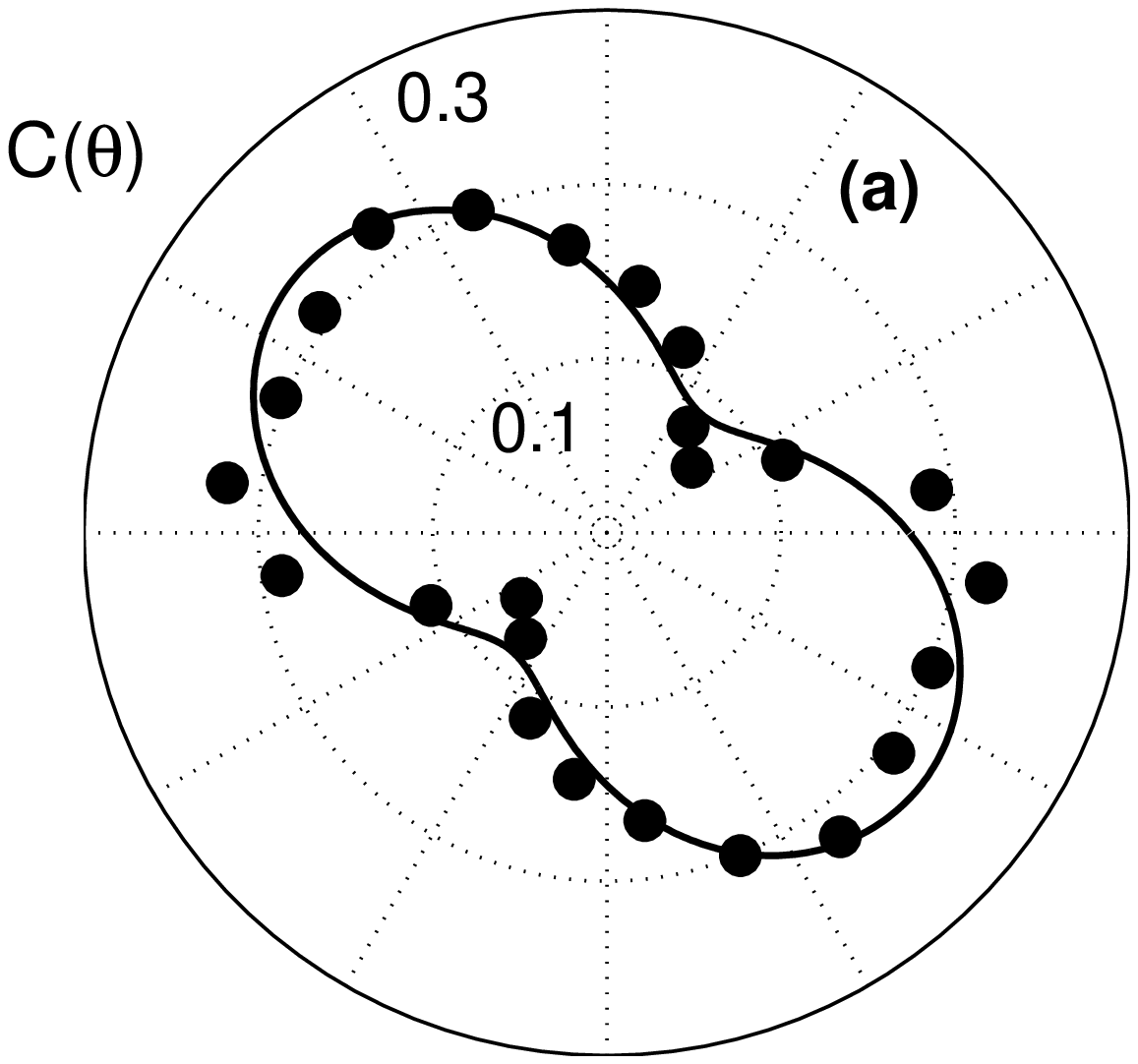}}
}
\mbox{
\scalebox{0.47}{\includegraphics{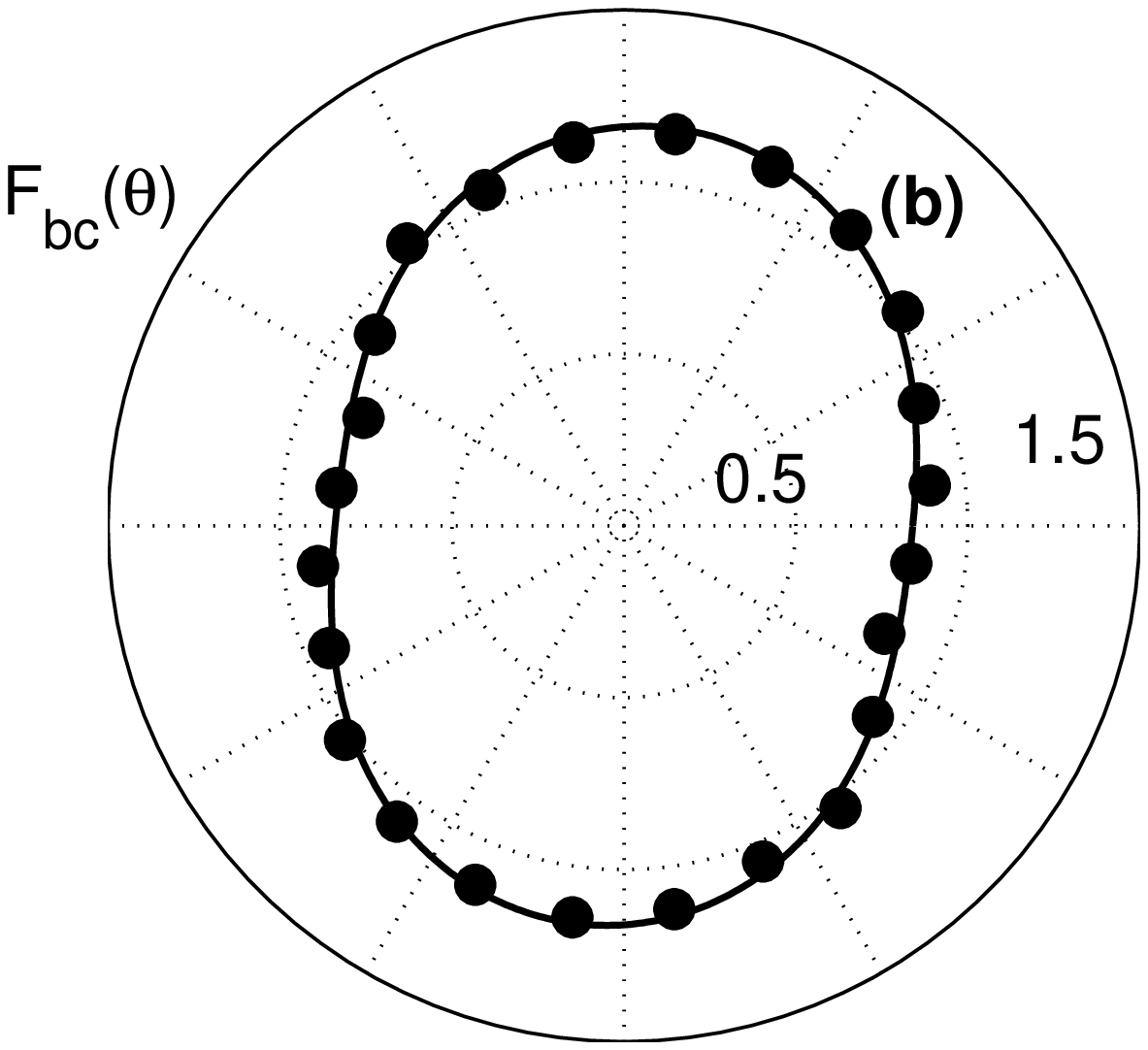}}
}
%\caption{\label{cfpolarplots} Polar plots of measurements (data points) and fits (lines) of {\bf (a)} the contact probability distribution $C(\theta)$ and {\bf (b)} the collisional force distribution $F_\mathrm{bc}(\theta)/\langle F_\mathrm{bc} \rangle$ for a granular material with $e=0$ and $\nu=0.79$.  The lines are fit to Equations~(\ref{angulareqns1}) and (\ref{angulareqns}), and the values of the Fourier components are plotted in Figure~\ref{acaffigure}.   
\caption{\label{cfpolarplots} Polar plots of measurements (data points) and fits (lines) of (a) the contact probability distribution $C(\theta)$ and (b) the collisional force distribution $F_\mathrm{bc}(\theta)/\langle F_\mathrm{bc} \rangle$ for a granular material with $e=0$ and $\nu=0.79$.  The lines are fits of Equations~(\ref{angulareqns1}) and (\ref{angulareqns}), and the values of the Fourier components are plotted in Figure~\ref{acaffigure}.   
}
\end{center}
\end{figure}

Equations~(\ref{appyo1}) and (\ref{constrainttwo}), combined with the basic force network Equation~(\ref{forceproptheta}), comprise the integral form of the force network model.  In order to carry out the integrations, it is necessary to know the functional form of $C(\theta)$ and $F_\mathrm{bc}(\theta)$.  These functions are $\pi$-periodic and can be written as Fourier Series, keeping only terms that are also $\pi$-periodic.  Previous research on the contact probability~\cite{thornton,arthur,rothenburg,anisradjai,kruyt} has shown that $C(\theta)$ is well approximated by keeping only the lowest Fourier terms.  We find that $F_\mathrm{bc}(\theta)$ has the same property.  We therefore approximate
\begin{eqnarray}
C(\theta) &=& \frac{1}{2 \pi} (1+a_c \sin{2 \theta} + a'_c \cos{2 \theta}),
\label{angulareqns1}
\\
F_\mathrm{bc}(\theta) &=& \langle F_\mathrm{bc} \rangle (1+a_f \sin{2 \theta} + a'_f \cos{2\theta}).
\label{angulareqns}
\end{eqnarray}
In Figure~\ref{cfpolarplots} we plot data of these functions for a granular material with $e=0$ and $\nu=0.79$, along with a fit to the above equations.  The fit is constructed by computing the fabric tensor $\phi_{\alpha \beta} = \langle \hat\sigma_\alpha \hat\sigma_\beta \rangle$ and force-fabric tensor $\chi_{\alpha \beta}^{(n)} =  \langle F_\mathrm{bc} \hat\sigma_\alpha \hat\sigma_\beta \rangle/\langle F_\mathrm{bc} \rangle$, where $\hat\sigma_\alpha$ is the $\alpha$ component of the unit vector between contacting grain centers, $F_\mathrm{bc}$ is the collisional force on the contact, and the average is taken over all contacts.  The anisotropies in the contact and force distributions $\{a_c,a'_c,a_f,a'_f\}$ are related to eigenvalues of the fabric tensors~\cite{anisradjai}, which are simple to measure.  We see from the plots that this first-order approximation for the contact probability and collisional force is quite good. 

\begin{figure}
\psfrag{aclabel}{\Huge{$a_c$}}
\psfrag{acprimelabel}{\Huge{$a'_c$}}
\psfrag{aflabel}{\Huge{$a_f$}}
\psfrag{afprimelabel}{\Huge{$a'_f$}}
\resizebox{!}{.48\textwidth}{{\includegraphics{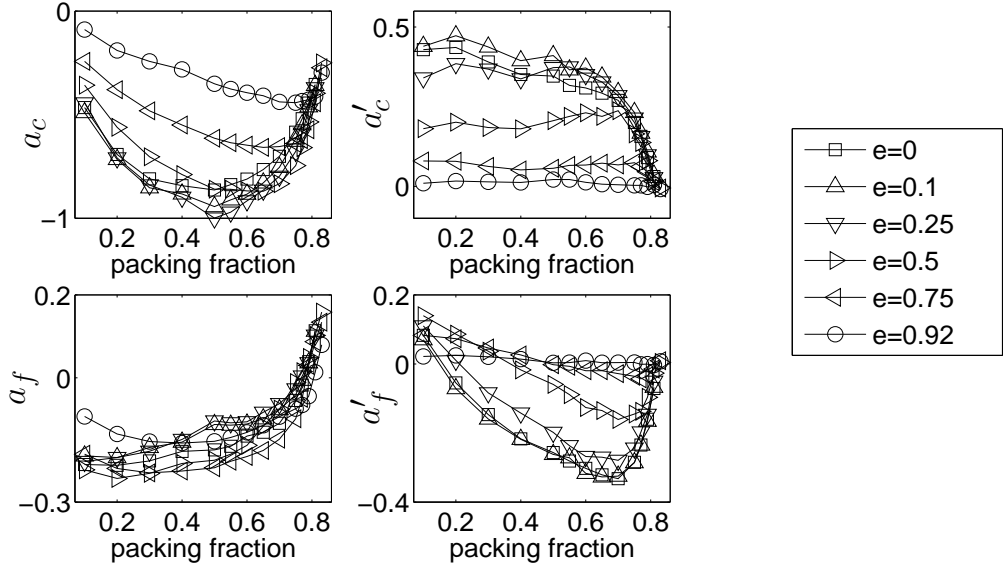}}}
\caption{\label{acaffigure} Measured values of the Fourier components from Equation~(\ref{angulareqns}).
}
\end{figure}

We have measured $C(\theta)$ and $F_\mathrm{bc}(\theta)$ for a wide range of restitution coefficients and packing fractions.  In Figure~\ref{acaffigure} we plot the value of the Fourier components from Equations~(\ref{angulareqns1}) and (\ref{angulareqns}), which characterize the functional form in all cases.  These plots reveal that the anisotropy in both the contact probability and collisional force depends sensitively on the value of the packing fraction and restitution coefficient.  The size of the components is of order $10^{-1}$ whereas the magnitude of the next order coefficients in the Fourier Series is much smaller.  This allows us to truncate the series in Equations~(\ref{angulareqns1}) and (\ref{angulareqns}) at first order and still get good agreement to the actual data, as in Figure~\ref{cfpolarplots}.
 
Now that we have a functional form for $C(\theta)$ and $F_\mathrm{bc}(\theta)$, we can solve for $\mathcal{F}_\ell(\theta)$ to first order in the Fourier components $\{a_c, a'_c, a_f, a'_f\}$.  This gives
\begin{equation}
\mathcal{F}_\ell(\theta) = \langle F_\mathrm{bc} \rangle (z-1)^\ell \Big( \Phi^\ell +\Psi^\ell(a_f \sin{2\theta}+a'_f \cos{2\theta}) +\sum_{i=1}^\ell \Psi^i \Phi^{\ell-i} (a_c \sin{2 \theta}+a'_c \cos{2\theta}) \Big), 
\label{prelimscriptf}
\end{equation}
where $\Phi$ and $\Psi$ are variables that depend on the geometry of the force networks and are expressible as 
\begin{equation}
\label{phipsi}
\{\Phi,\Psi\} = \int_{-2 \pi/3}^{2 \pi/3} dx P_C(x) \cos(x) \{1,\cos(2 x) \}.
\end{equation}
We can also solve for the constraint in Equation~(\ref{constrainttwo}).  To lowest order in the Fourier components, the constraint equation gives
\begin{equation} 
\frac{T(\theta)}{\langle F_\mathrm{bc}(\theta)\rangle} = \Phi (z-1) \leq 1.
\end{equation} 
This, combined with Equation~(\ref{prelimscriptf}), provides a closed formula for $\mathcal{F}_\ell$:
\begin{eqnarray} 
\mathcal{F}_\ell(\theta) = \langle F_\mathrm{bc} \rangle &\Bigg(& \min[\Phi(z-1),1]^\ell +\Psi^\ell(z-1)^\ell(a_f \sin{2\theta}+a'_f \cos{2\theta}) \\
&+&\sum_{i=1}^\ell \Psi^i (z-1)^i \min[\Phi(z-1),1]^{\ell-i} (a_c \sin{2 \theta}+a'_c \cos{2\theta}) \,\, \Bigg), 
\label{finalscriptf}
\end{eqnarray} 
to first order in $\{a_c,a'_c,a_f,a'_f\}$.

This solution can now be used to arrive at a constitutive relation for the stress tensor.  The static stress tensor is given by Equation~(\ref{sstress}), which can be rewritten in two-dimensions as 
\begin{equation}
\Sigma^\mathrm{s}_{\alpha \beta} = \frac{1}{V} \int d\theta C(\theta) \sigma(\theta) F(\theta) \times 
\label{integralsstress}
\begin{pmatrix}
\cos^2\theta & \cos\theta \sin\theta \\ \cos\theta \sin\theta & \sin^2\theta 
\end{pmatrix}, 
\end{equation}
where $\sigma(\theta)$ is the average value of the distance between grains at contact for a given angle.  In our simulations we observe that $\sigma(\theta)$ has very little dependence on $\theta$ (of order less than $10^{-4}$).  Thus $\sigma(\theta) = \langle \sigma \rangle$.  We can also use this same integral form to determine the collisional stress tensor by replacing $F(\theta)$ with $F_\mathrm{bc}(\theta)$.  

In this paper we have concentrated on the pressure and shear stress.  The pressure is given by one-half the trace of Equation~(\ref{integralsstress}) and the shear stress by either off-diagonal element, but these two quantities do not fully describe the stress tensor.  There is a third independent term and, without loss of generality, we use $\Sigma_{11}$.  Inserting the solution for $F(\theta)$ from Equations~(\ref{forceproptheta}) and (\ref{finalscriptf}) into Equation~(\ref{integralsstress}), we arrive at the following constitutive relations that fully describe the stress tensor:
\begin{eqnarray}
\label{pconst}
\frac{p^\mathrm{s}-p^\mathrm{bc}}{p^\mathrm{bc}} &=& \sum_{\ell=1}^{\xi/\xi_\mathrm{el}-1} \min[\Phi (z-1),1]^\ell, \\
\label{sconst}
\frac{s^\mathrm{s}-s^\mathrm{bc}}{p^\mathrm{bc}} &=& \frac{1}{2} \sum_{\ell=1}^{\xi/\xi_\mathrm{el}-1} \Bigg( a_f \Psi^\ell (z-1)^\ell + a_c \sum_{i=0}^{\ell} \Psi^i (z-1)^i \min[\Phi (z-1),1]^{\ell-i} \Bigg) , \\
\frac{\Sigma^\mathrm{s}_{11} - \Sigma^\mathrm{bc}_{11}}{p^\mathrm{bc}} &=& \sum_{\ell=1}^{\xi/\xi_\mathrm{el}-1} \Bigg( \min[\Phi (z-1),1]^\ell + \frac{a'_f}{2} \Psi^\ell (z-1)^\ell \nonumber \\
&+& \frac{a'_c}{2} \sum_{i=0}^{\ell} \Psi^i (z-1)^i \min[\Phi (z-1),1]^{\ell-i} \Bigg).
\label{sxxconst}
\end{eqnarray}
These equations relate the static stress tensor to its collisional values, and properties of the force networks.  
The left hand side (lhs) of each equation gives the difference between the static and collisional values of stress.  These differences are equal to network properties on the right hand side (rhs) of each equation, which are summed over all possible force chain path lengths $\ell$ in the network.  A large number of network parameters appear in these equations and makes the form rather complicated.  This is to be expected, since the structure of the force networks is complex and the predictions of Equations~(\ref{pconst}-\ref{sxxconst}) span system behavior from very dilute granular materials with $\phi \approx 0$ to dense granular materials with $\phi$ arbitrarily close to $\phi_\mathrm{c}$ (and for all restitution coefficients $e$).  The properties of the networks change considerably over this range of parameter space, and they must be retained in the Equations.             

Simple scaling relations can be obtained near certain packing fractions.  For example, near the network transition at $\nu_\mathrm{bc}$, Equations~(\ref{pconst}-\ref{sxxconst}) predict that $(\Sigma^\mathrm{s}_{\alpha \beta}-\Sigma^\mathrm{bc}_{\alpha\beta}) \propto (z-1)$.  This is because, when $\xi/\xi_\mathrm{el} \approx 1$ near $\nu_\mathrm{bc}$, the deviation from the collisional stress is dominated by forces transmitted between nearest neighbors.  Therefore the excess number of contacts serves as the dominant scaling variable.  In contrast, near the jamming transition at $\nu_\mathrm{c}$, networks are saturated so that Equation~(\ref{constraintone}) takes its maximum value and all of the kinetic energy from each contact is transferred to the network.  In this case the stress tensor should depend on the size of the networks, and the constitutive relations indeed predict that $\Sigma^\mathrm{s}_{\alpha \beta} \propto \xi$.  This scaling is especially interesting since it suggests that the size of the networks is the important scaling variable, which might also control other features of the jamming transition.  Indeed, once $\xi$ becomes large, it is clusters of grains, and not individual grains, that serve as the basic thermodynamic degrees of freedom. 

Finally, we consider the limit of $\xi/\xi_\mathrm{el} \rightarrow 1$ where force networks consist of only two grains.  In this limit the force network model predicts that $\Sigma^\mathrm{s}_{\alpha \beta} = \Sigma^\mathrm{bc}_{\alpha \beta}$, which is the appropriate result in the dilute regime where kinetic theory holds.  

%They only encompass two of the three independent values of the stress tensor.  It is also possible to derive a constitutive relation for the first normal stress difference given by $(\Sigma_{xx}-\Sigma_{yy})/2$.  Indeed the equation for this quantity is exactly equivalent to Equation~(\ref{sconst}) with $a_c$ and $a_f$ replaced by $a'_c$ and $a'_f$. 

\subsection{Testing the predictions}
\label{testsection}
Equations~(\ref{pconst}-\ref{sxxconst}) make predictions for all independent components of the stress tensor over the complete range of $e$ and for $\phi<\phi_\mathrm{c}$.  The difference between the static and collisional values of stress is related to features of the force networks.  These include the anisotropies in the contact probability and collisional force $\{a_c,a'_c,a_f,a'_f\}$, the size of the force networks $\xi/\xi_\mathrm{el}$, the average coordination number $z$, and a pair of geometric variables $\{\Phi,\Psi\}$ that are defined in Equation~(\ref{phipsi}) and are related to the distribution of contacts on single grains.  All of these variables except $z$, $\Phi$, and $\Psi$ have been measured previously in this paper or in Ref.~\cite{gajcompanion1}.  Our next step is to measure $z$, $\Phi$, and $\Psi$.

\begin{figure}
\begin{center}
\mbox{
\psfrag{yl}{\Huge{${z}$}}
\psfrag{iyl}{\Huge{${z}$}}
\scalebox{0.32}{\includegraphics{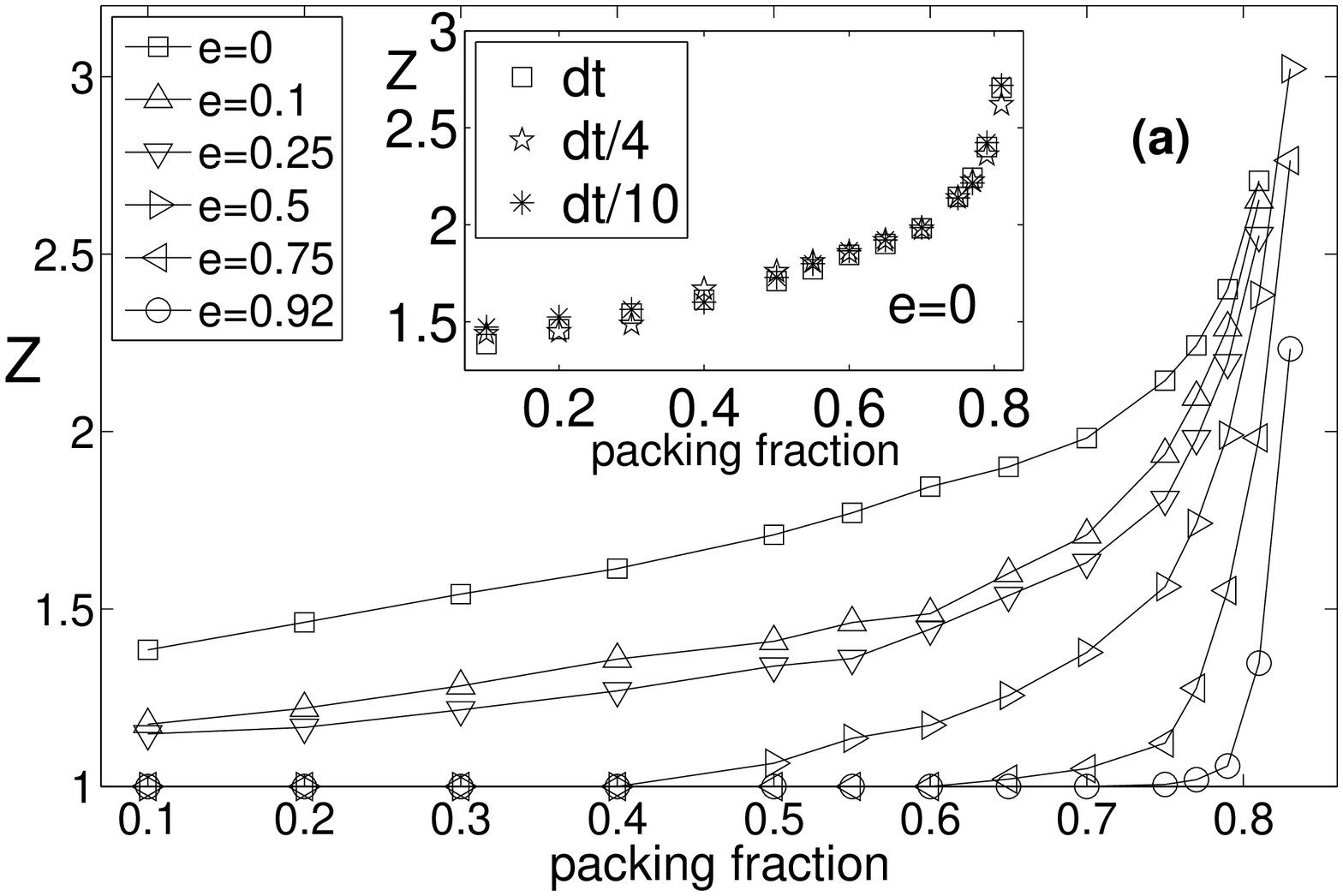}}
}
\mbox{
\psfrag{uyl }{\Huge{$\Phi$}}
\psfrag{lyl }{\Huge{$\Psi$}}
\scalebox{0.32}{\includegraphics{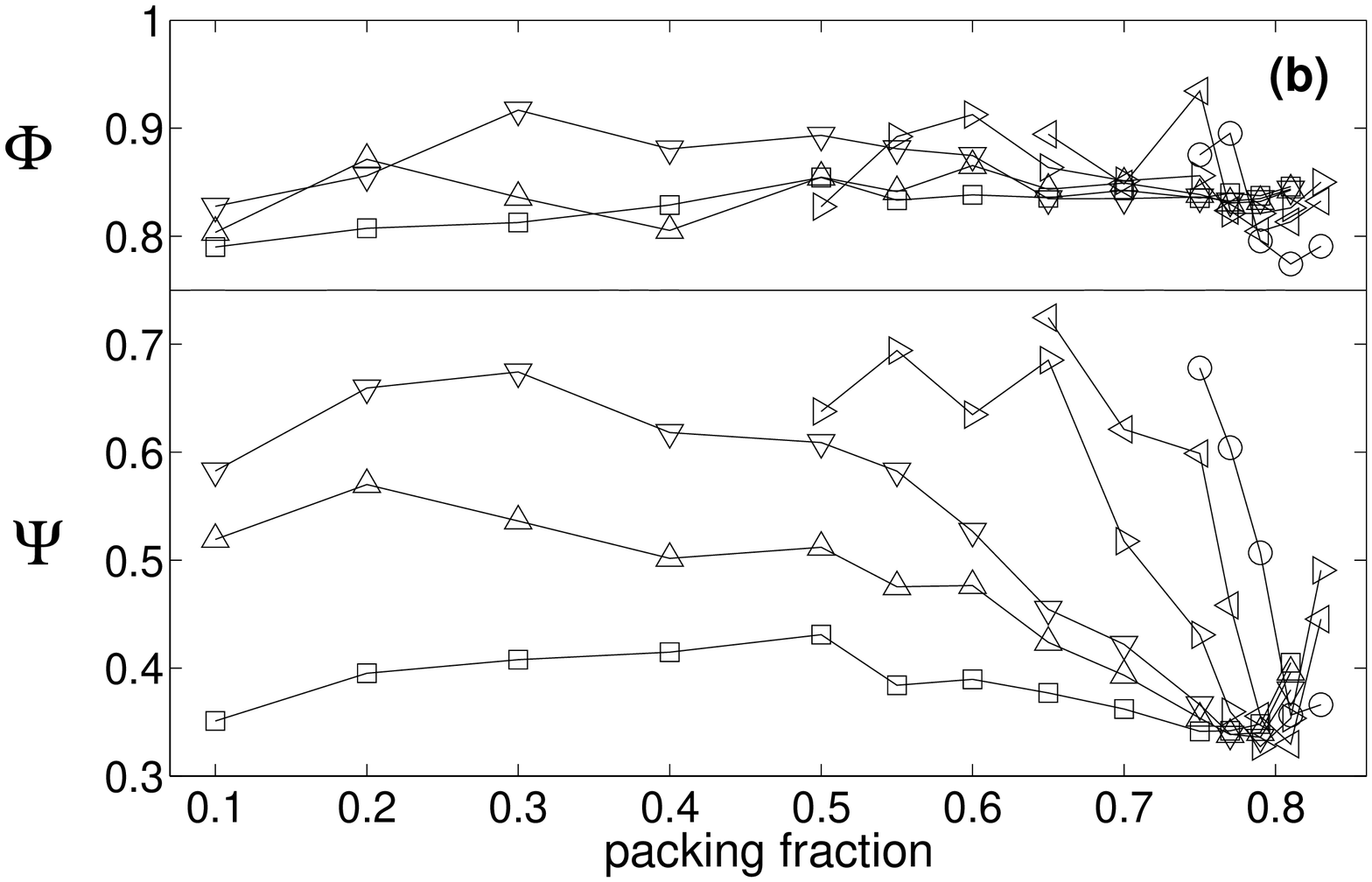}}
}
%\caption{\label{networkprops} {\bf (a)} The coordination number $z$ and {\bf (b)} $\Phi$ and $\Psi$ (Equation~(\ref{phipsi})) for a wide variety of packing fractions and restitution coefficients.  The line labels in {\bf (b)} are the same as in {\bf (a)}.  Although $z$ and $\Psi$ depend sensitively on the value of $\nu$ and $e$, the value of $\Phi$ is always approximately $0.83$.
\caption{\label{networkprops} (a) The coordination number $z$ and (b) $\Phi$ and $\Psi$ (Equation~(\ref{phipsi})) for a wide variety of packing fractions and restitution coefficients.  The line labels ($e$-values) in (b) are the same as in (a).  Although $z$ and $\Psi$ depend sensitively on the value of $\nu$ and $e$, the value of $\Phi$ is always approximately $0.83$.
}
\end{center}
\end{figure}

In Figure~\ref{networkprops} we plot the values of $z$, $\Phi$, and $\Psi$ as measured in our simulations.  We measure $z$ by averaging over long-lived contacts, which are pairs of contacting grains that were also contacting in the previous time step.  This ensures that only the static backbone of the force network is considered and that transient contacts do not artificially increase the coordination number.  This measurement does not depend on the time step, as shown in the inset.  We also measure $\Phi$ and $\Psi$, as prescribed in Equations~(\ref{definepc}) and (\ref{phipsi}), by averaging over the same set of contacts.  We observe that $\Psi < \Phi$ for all granular materials we have considered.      

\begin{figure}
\begin{center}
\mbox{
%\psfrag{lhs}{\Huge{$\frac{p^\mathrm{s}-p^\mathrm{bc}}{p^\mathrm{bc}}$}}
\scalebox{0.37}{\includegraphics{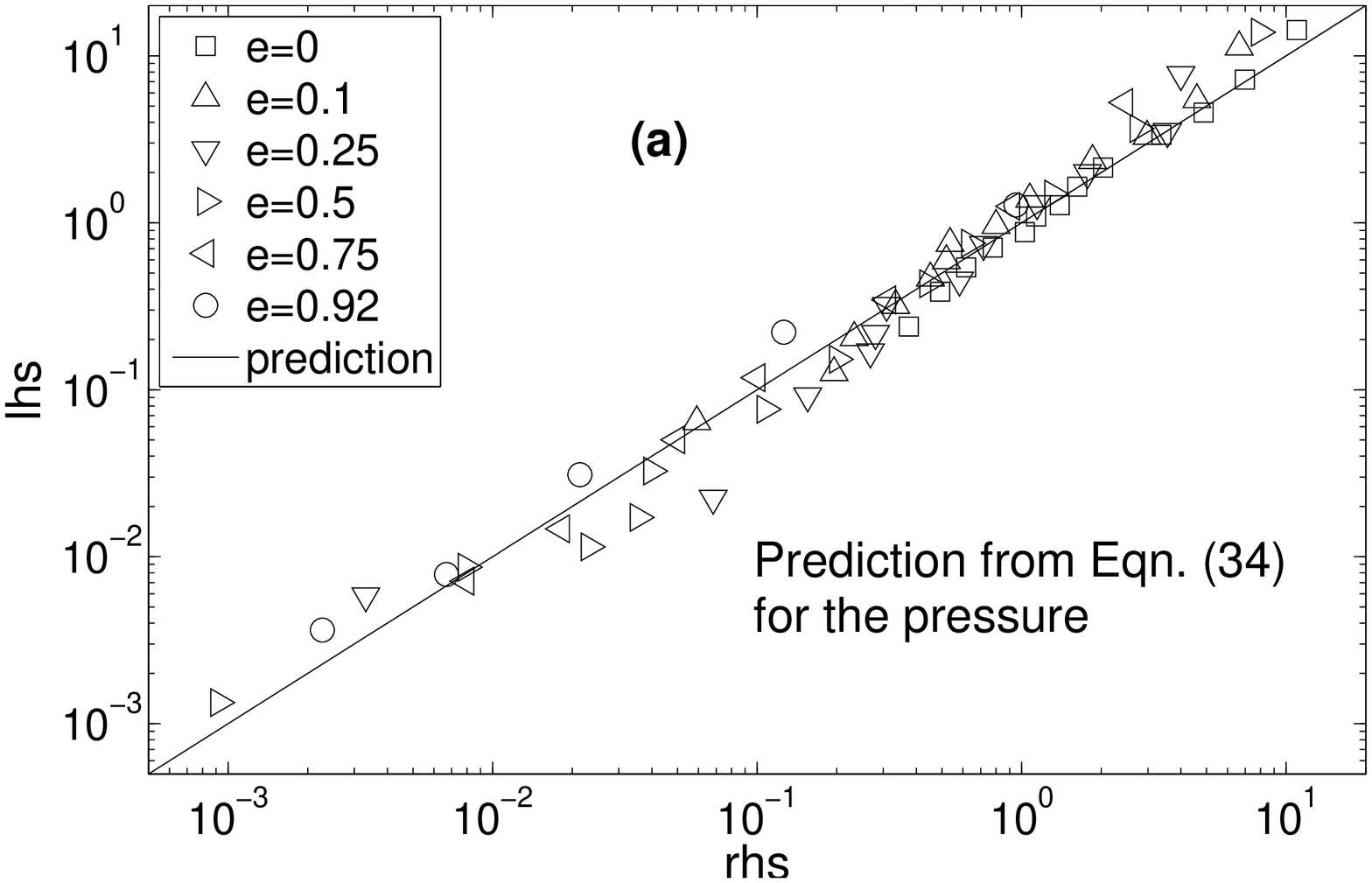}}
}
\mbox{
%\psfrag{lhs}{\Huge{$\frac{s^\mathrm{s}-s^\mathrm{bc}}{p^\mathrm{bc}}$}}
\scalebox{0.37}{\includegraphics{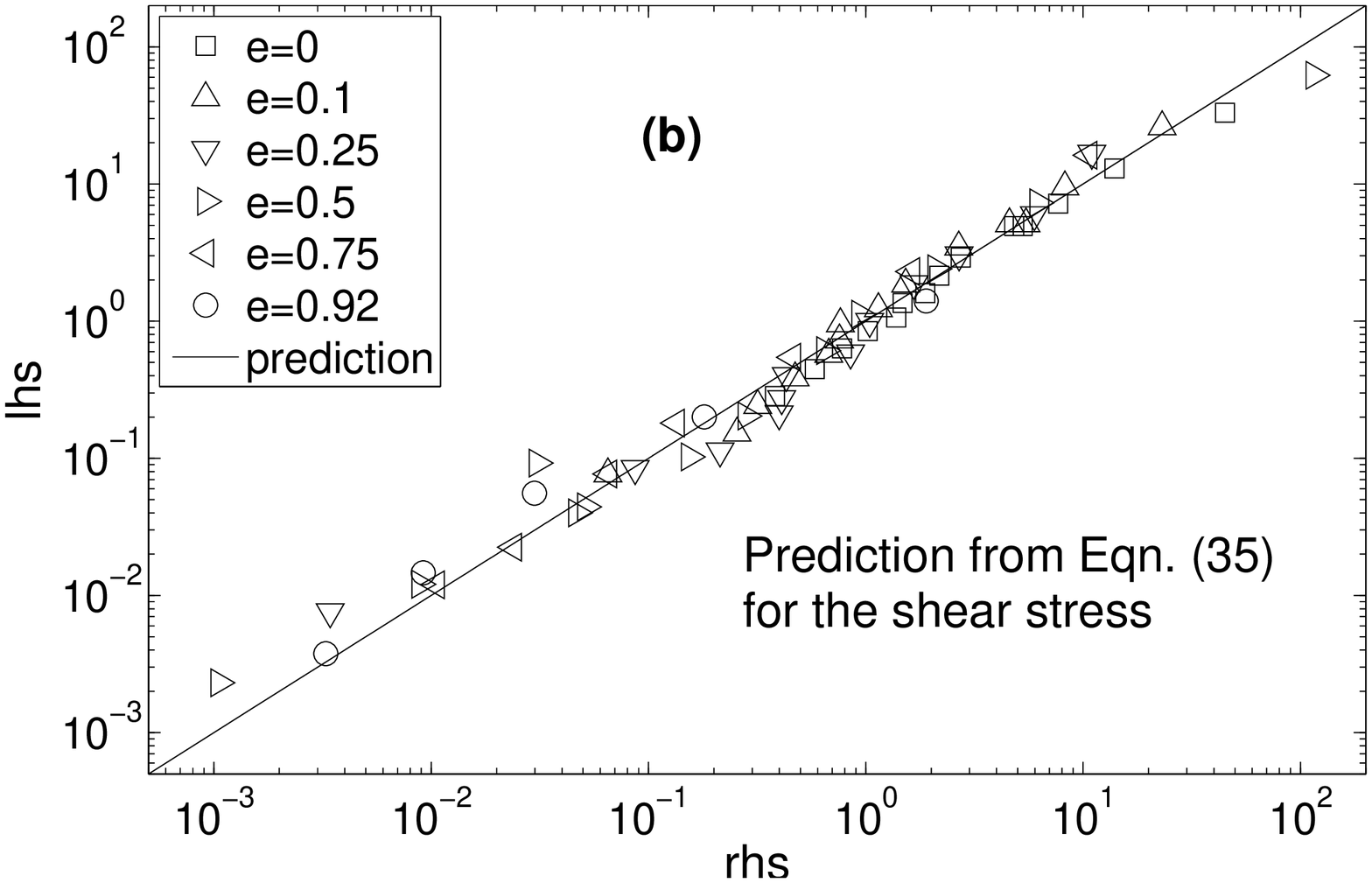}}
}
\mbox{
%\psfrag{lhs}{\Huge{$\frac{\Sigma_{11}^\mathrm{s}-\Sigma_{11}^\mathrm{bc}}{p^\mathrm{bc}}$}}
\scalebox{0.37}{\includegraphics{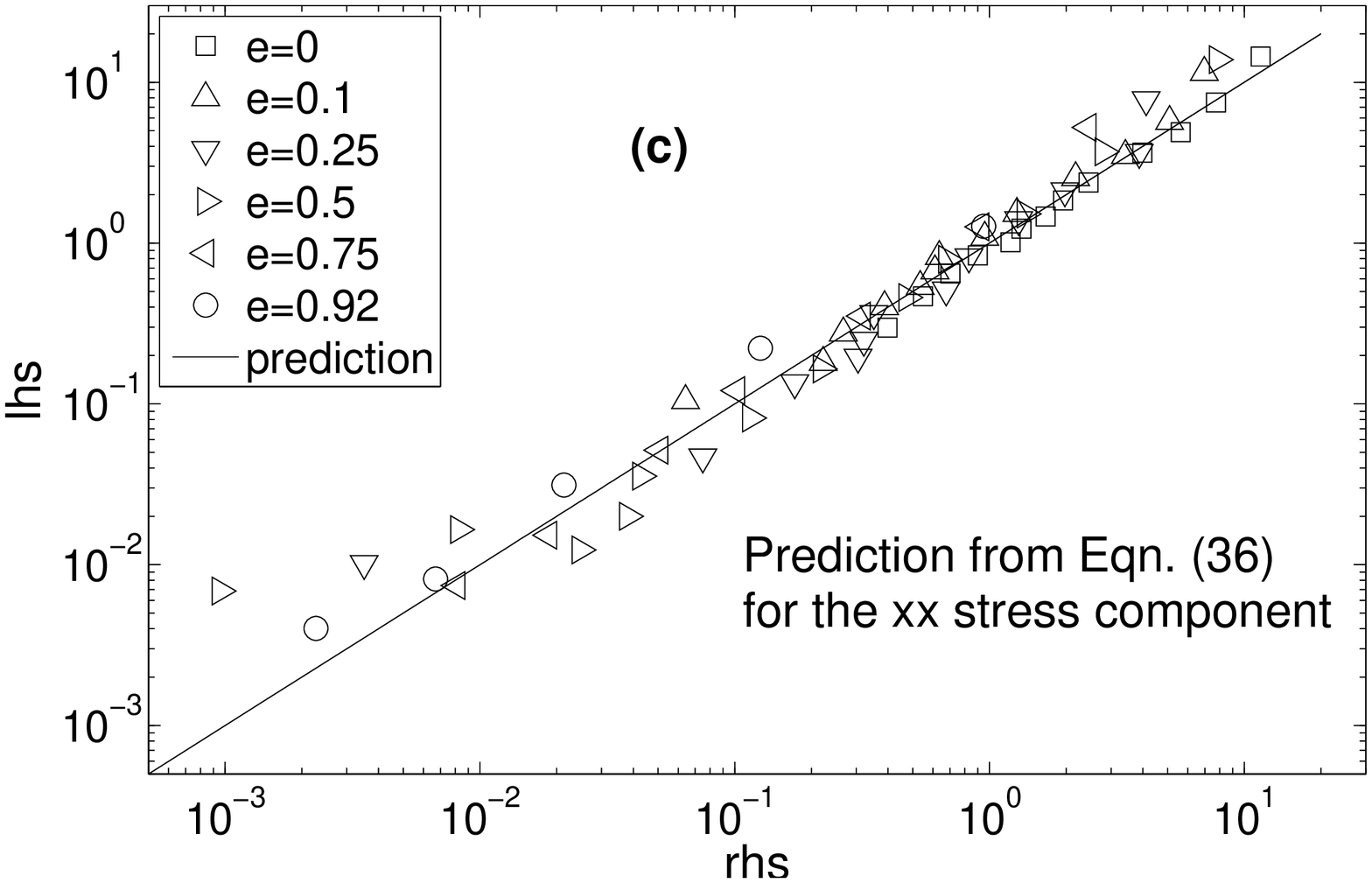}}
}
%\caption{\label{consttest} Tests of the constitutive relations from the force network model.  {\bf (a)} Test of Equation~(\ref{pconst}).  The left hand side (lhs) of the equation, $\frac{p^\mathrm{s}-p^\mathrm{bc}}{p^\mathrm{bc}}$, is plotted as a function of the right hand side (rhs).  This collapses the data to the line predicted by the model; {\bf (b)} The left hand side of Equation~(\ref{sconst}), $\frac{s^\mathrm{s}-s^\mathrm{bc}}{p^\mathrm{bc}}$, plotted as a function of the right hand side, once again collapsing to the prediction;  {\bf (c)}  The left hand side of Equation~(\ref{sxxconst}), $\frac{\Sigma_{11}^\mathrm{s}-\Sigma_{11}^\mathrm{bc}}{p^\mathrm{bc}}$, plotted as a function of the right hand side.  All of the plots have been constructed using simulation data for each variable and no fitting parameters have been utilized.
\caption{\label{consttest} Tests of the constitutive relations from the force network model.  (a) Test of Equation~(\ref{pconst}).  The left hand side (lhs) of the equation, $\frac{p^\mathrm{s}-p^\mathrm{bc}}{p^\mathrm{bc}}$, is plotted as a function of the right hand side (rhs).  This collapses the data to the line predicted by the model; (b) The left hand side of Equation~(\ref{sconst}), $\frac{s^\mathrm{s}-s^\mathrm{bc}}{p^\mathrm{bc}}$, plotted as a function of the right hand side, once again collapsing to the prediction;  (c)  The left hand side of Equation~(\ref{sxxconst}), $\frac{\Sigma_{11}^\mathrm{s}-\Sigma_{11}^\mathrm{bc}}{p^\mathrm{bc}}$, plotted as a function of the right hand side.  All of the plots have been constructed using simulation data for each variable and no fitting parameters have been utilized.
}
\end{center}
\end{figure}

We have now measured every variable in the constitutive relations of Equations~(\ref{pconst}-\ref{sxxconst}).  We can therefore test the validity of the predictions without using any fitting parameters.  Due to the complexity of the equations, it is convenient to plot the right hand side (rhs) of each equation versus the left hand side (lhs).  This is shown in Figure~\ref{consttest} using all of the data we have collected.  Plotted in this way, the data for each component of the stress tensor collapses onto the line predicted by the force network model over more than four decades.  This collapse is especially striking since the variables in the predictions have a wide variance as a function of both restitution coefficient and packing fraction.

The collapse of our data onto the predicted curves suggests that the force network model captures an essential property of granular materials over a broad range of densities and restitution coefficients.  The success of the model is based on visualizing granular materials as conglomerates of interacting networks, instead of collections of grains.  Thus the packing fraction and restitution coefficient, which are grain properties, are substituted by the size, coordination, and other properties of the networks.  This allows for constitutive relations to be determined analytically.  
%To make connection with experiment, where the network properties are often difficult to measure, it is necessary to determine the relationship between grain properties and network properties.  This step will enable the stress tensor to be derived directly from grain properties, and unambiguously predict the stress tensor in experimental settings.  

Finally, it is important to remark that the constitutive relations from the force network model have been derived in the limit of perfectly rigid grains, which may not always apply to realistic flows with finite grain stiffness.  In the case of finite grain stiffness, there is a finite speed $v_c$ at which forces propagate through the network.  Combined with the lifetime of the networks $\tau_c$, this sets a maximum correlation length $v_c \tau_c$, since information can only be transferred between a pair of grains if the network exists long enough to propagate it.  This maximum correlation length is a monotonically increasing function of the grain stiffness.  If $v_c \tau_c > \xi$, then the stress tensor can be described by Equations~(\ref{pconst}-\ref{sxxconst}).  However, if $v_c \tau_c < \xi$, it is necessary to replace the length scale $\xi$ with $v_c \tau_c$.  Because $\xi$ diverges as the material approaches the jamming limit and $v_c \tau_c$ is always finite, we expect that for a given grain stiffness there is a critical packing above which $v_c \tau_c < \xi$.  This critical packing fraction is always in the inertial regime and must be strictly less than $\nu_c$.  Therefore, very close to jamming, the elasticity of grains begins to play an important role.  For all other packing fractions, the assumption of perfectly rigid grains is valid for natural and experimental flows.

\section{Conclusions} 
The underlying microscopic interactions between grains have a large influence on macroscopic characteristics of granular flow.  We have investigated two models of the stress tensor-- kinetic theory and the force network model-- which make different assumptions about the microscopic interactions.  Kinetic theory assumes that only binary collisions are relevant and calculates the stress tensor based on grain properties, whereas the force network model allows simultaneous interaction between many grains and calculates the stress tensor based on properties of the resultant force networks.    

For dilute flows, which occur when the size of the force networks is small, kinetic theory makes accurate predictions.  This is not surprising since small force networks imply localized interactions and binary collisions.  For dense flows, force networks extend beyond pairs of grains and the predictions of kinetic theory no longer match data from simulations.  This is because grain-grain correlations are induced via the force networks and kinetic theory does not take them into account.  However, correlations never exist between isolated networks, and by constructing the force network model based on network properties, we are able to accurately predict the stress tensor for both dilute and dense flows.
%In the dense regime the .  Of course, the force network model also applies in the dilute regime, where the predictions match those of kinetic theory.        

The force network model predicts all independent components of the stress tensor over the entire inertial regime and matches data from simulations for more than four decades.  
Extensions of the model could be used to predict other quantities, including the contact force distribution function $P(f)$.  An integral part of any such theory is the finite size of the force networks, which has important effects on the qualitative features of $P(f)$~\cite{gajcompanion1}.  
Further extensions could also specify relations between network parameters and thereby simplify the constitutive equations~(\ref{pconst}-\ref{sxxconst}).  While it is possible that the relations between network parameters are complicated and depend on many factors, it is likely that simple scalings exist near the network transition at $\nu_\mathrm{bc}$.  For $\nu \approx \nu_\mathrm{bc}$ the deviation between the static and collisional stress tensors is proportional to $(z-1)$ and it is likely that network parameters also scale with powers of $(z-1)$.  In particular, it would be interesting to probe the dependence of $\xi$ on $(z-1)$, although fluctuations in the parameters have complicated our measurement of this relation.

Finally, while we have concentrated on the inertial regime, it is also important to understand how natural flows make the transition from dynamics dominated by inertia, to quasi-static dynamics, and ultimately to how the system jams.  Along this sequence, the stiffness of the grains plays an increasingly active role in the dynamics.  The force network model can accommodate the development of stiffness by incorporating a maximum length scale through which forces can propagate.  Including this mechanism for a granular material moving through $\nu_c$ may help connect the dense inertial regime with the quasi-static regime and facilitate a more complete understanding of dry granular materials. 

This work was supported by the William M. Keck Foundation, the MRSEC program of NSF under Award No. DMR00-80034, the James S. McDonnell Foundation, the David and Lucile Packard Foundation, and NSF Grant Nos. DMR-9813752, PHY99-07949 and DMR-0606092.


\begin{thebibliography}{100}
\bibitem{aransonreview}
I. S. Aranson and L. S. Tsimring, Rev. Mod. Phys. {\bf 78}, 641 (2006).

\bibitem{gajcompanion1}
G. Lois, A. Lemaitre, and J. M. Carlson, preceding paper

\bibitem{qmodel}
S. N. Coppersmith, C. -h. Liu, S. Majumdar, O. Narayan, and T. A. Witten, Phys. Rev. E {\bf 53}, 4673 (1996).

\bibitem{socolaralpha}
J. E. S. Socolar, Phys. Rev. E {\bf 57}, 3204 (1998).

\bibitem{claudinstress}
P. Claudin, J.-P. Bouchaud, M. E. Cates, and J. P. Wittmer, Phys. Rev. E {\bf 57}, 4441 (1998).

\bibitem{nicodemi}
M. Nicodemi, Phys. Rev. Lett. {\bf 80}, 1340 (1998).

\bibitem{socolarforce}
J. E. S. Socolar, D. G. Schaeffer, and P. Claudin, Eur. Phys. J. E {\bf 7}, 353 (2002).

\bibitem{ottoforce}
M. Otto, J.-P. Bouchaud, P. Claudin, and J. E. S. Socolar, Phys. Rev. E {\bf 67}, 031302 (2003).

\bibitem{bouchaudforce}
J.-P. Bouchaud, P. Claudin, D. Levine and M. Otto,  Eur. Phys. J. E {\bf 4}, 451 (2001).

\bibitem{ogawa}
S. Ogawa, {\em Proc. US-Jpn. Semin. Contin.-Mech. and Stat. Approaches Mech. Granular Mater.}, pp. 208-217 (Tokyo, Gukujutsu Bunken, 1978).

\bibitem{haff}
P. K. Haff, J. Fluid Mech. {\bf 134}, 401 (1983).

\bibitem{jenkinssavage}
J. T. Jenkins and S. B. Savage, J. Fluid Mech. {\bf 130}, 87 (1983).

\bibitem{lun}
C. K. K. Lun, S. B. Savage, D. J. Jeffrey, and N. Chepurniy, J. Fluid Mech. {\bf 140},223 (1984).

\bibitem{jenkins}
J. T. Jenkins and M. W. Richman, Phys. Fluids {\bf 28}, 3485 (1985).

\bibitem{garzodufty}
V. Garzo and J. W. Dufty, Phys. Rev. E {\bf 59}, 5895 (1999).

\bibitem{lutsko1}
J. F. Lutsko, Phys. Rev. E {\bf 70}, 061101 (2004).

\bibitem{lutsko2}
J. F. Lutsko, Phys. Rev. E {\bf 72}, 021306 (2005).

\bibitem{campbellreview}
C. S. Campbell, Annu. Rev. Fluid Mech. {\bf 22}, 57 (1990).

\bibitem{goldhirschreview}
I. Goldhirsch, Annu. Rev. Fluid Mech. {\bf 35}, 267 (2003).

\bibitem{vannoijebook}
T. P. C. van Noije and M. H. Ernst in {\em Lecture Notes in Physics} {\bf 564}, edited by T. Poschel and S. Luding (Springer-Verlag 2001).

\bibitem{azanza}
E. Azanza, F. Chevoir, and P. Moucheront, J. Fluid. Mech. {\bf 400}, 199 (1999).

\bibitem{zhang1}
D. Z. Zhang and R. M. Rauenzahn, J. Rheol. {\bf 44}, 1019 (2000).

\bibitem{shen}
H. H. Shen and B. Sankaran, Phys. Rev. E {\bf 70}, 051308 (2004).

\bibitem{zhang2}
D. Z. Zhang, Phys. Rev. E {\bf 71}, 041303 (2005).

\bibitem{gaj2}
G. Lois, A. Lemaitre, and J. M. Carlson, Europhys. Lett. {\bf 76}, 318 (2006).

\bibitem{gaj3}
G. Lois, A. Lemaitre, and J. M. Carlson, to appear, Computers and Mathematics (2006).

\bibitem{breystat}
J. J. Brey, J. W. Dufty, and A. Santos, J. Stat. Phys. {\bf 87}, 1051 (1997).

\bibitem{ktbook1}
S. Chapman and T. G. Cowling, {\em The Mathematical theory of non-uniform gases} (Cambridge, 1970).

\bibitem{ktbook2}
J. H. Ferziger and H. G. Kaper, {\em Mathematical theory of transport processes in gases} (Elsevier, New York 1972).

%\bibitem{johnsonjackson}
%P. C. Johnson and R. Jackson, J. Fluid Mech. {\bf 176}, 67 (1987).
%
%\bibitem{aranson1}
%I. S. Aranson and L. S. Tsimring, Phys. Rev. E {\bf 65}, 061303 (2002).
%
%\bibitem{aranson2}
%D. Volfson, L. S. Tsimring, and I. S. Aranson, Phys. Rev. E {\bf 68}, 021301 (2003).
%
%\bibitem{goldhirschcluster}
%I. Goldhirsch and G. Zanetti, Phys. Rev. Lett. {\bf 70}, 1619 (1993).
%
%\bibitem{mcnamaracluster}
%S. McNamara and W. R. Young, Phys. Rev. E {\bf 53}, 5089 (1996).
%
%\bibitem{hrenyacollapse}
%M. Alam and C. M. Hrenya, Phys. Rev. E {\bf 63}, 061308 (2001).
%
%\bibitem{halsey}
%T. C. Halsey and D. Ertas, preprint, cond-mat/0506170.

\bibitem{lusa}
S. Luding and A. Santos, J. Chem. Phys. {\bf 121}, 8458 (2004).

\bibitem{shattuck}
S. J. Moon, M. D. Shattuck, J. B. Swift, Phys. Rev. E {\bf 64}, 031303 (2001).

\bibitem{vannoijepre}
T. P. C. van Noije and M. H. Ernst, preprint, cond-mat/9706020.

\bibitem{mesoscopictheory}
T. P. C. van Noije, M. H. Ernst, R. Brito, and J. A. G. Orza, Phys. Rev. Lett. {\bf 79}, 411 (1997).

\bibitem{blairkudrolli}
D. L. Blair and A. Kudrolli, Phys. Rev. E {\bf 64}, 050301(R) (2001).

\bibitem{prevost}
A. Prevost, D. A. Egolf, and J. S. Urbach, Phys. Rev. Lett. {\bf 89}, 084301 (2002).

\bibitem{pouliquenvelcorr}
O. Pouliquen, Phys. Rev. Lett. {\bf 93}, 248001 (2004).

\bibitem{grestsilbert}
L. E. Silbert, D. Ertas, G. S. Grest, T. C. Halsey, and D. Levine, Phys. Rev. E {\bf 65}, 051302 (2002).

\bibitem{dacruz}
F. da Cruz, S. Emam, M. Prochnow, J.-N. Roux, and F. Chevoir, Phys. Rev. E {\bf 72}, 021309 (2005).

\bibitem{gaj1}
G. Lois, A. Lemaitre, and J. M. Carlson, Phys. Rev. E {\bf 72}, 051303 (2005).

\bibitem{thornton}
C. Thornton and D. J. Barnes, Acta Mechanica {\bf 64}, 45 (1986).

\bibitem{arthur}
J. R. F. Arthur, J. A. Koenders, and R. K. S. Wong, Acta Mechanica {\bf 64}, 19 (1986).

\bibitem{rothenburg}
L. Rothenburg and R. J. Bathurst, Geotechnique {\bf 39}, 601 (1989).

\bibitem{anisradjai}
F. Radjai, D. E. Wolf, M. Jean, and J.-J. Moreau, Phys. Rev. Lett. {\bf 80}, 61 (1998).

\bibitem{kruyt}
N. P. Kruyt, Int. J. of Sol. and Struct., {\bf 40}, 3537 (2003).

\bibitem{radjai3}
H. Troadec, F. Radjai, S. Roux and J. C. Charmet, Phys. Rev. E {\bf 66}, 041305 (2002).

\end{thebibliography}
\end{document}